\documentclass[a4paper,12pt]{spieman}  % use this instead for A4 paper
\usepackage{aas_macros}
\usepackage{amsmath,amsfonts,amssymb}
\usepackage{graphicx}
\usepackage{setspace}
\usepackage{tocloft}
\usepackage{threeparttable} % for table footnotes
\usepackage{rotating} % for sideways tables
\usepackage{booktabs}
\usepackage{array}
\usepackage{lineno}
\usepackage{lscape}
\usepackage{color}

\title{Instrumentation for solar spectropolarimetry: State of the art and prospects}
\author[a,*]{F. A. Iglesias}
\author[b]{A. Feller}
\affil[a]{Universidad Tecnológica Nacional - Facultad Regional Mendoza, CONICET, CEDS, Rodruiguez 243, Mendoza, Argentina, 5500}
\affil[b]{Max Planck Institute for Solar System Research, Justus-von-Liebig-Weg 3, Göttingen, Germany, 37077}

\cftpagenumbersoff{figure}
\cftpagenumbersoff{table} 
\begin{document} 
\maketitle

\begin{abstract}
% 200 WORDS
Given its unchallenged capabilities in terms of sensitivity and spatial resolution, the combination of imaging spectropolarimetry and numeric Stokes inversion represent the dominant technique currently used to remotely sense the physical properties of the solar atmosphere, and in particular, its important driving magnetic field. Solar magnetism manifests itself in a wide range of spatial, temporal and energetic scales. The ubiquitous but relatively small and weak fields of the so called quiet Sun are believed today to be crucial for answering many open questions in solar physics, some of which have substantial practical relevance due to the strong Sun-Earth connection. However, such fields are very challenging to detect because they require spectropolarimetric measurements with high spatial (sub-arcsec), spectral ($<100$ mÅ) and temporal ($<10$ s) resolution along with high polarimetric sensitivity ($< 0.1$\% of the intensity). 
In this review we collect and discuss both well-established and upcoming instrumental solutions developed during the last decades to push solar observations towards the above-mentioned parameter regime. This typically involves design trade-offs due to the high dimensionality of the data and signal-to-noise-ratio considerations, among others. We focus on the main three components that form a spectropolarimeter, namely, wavelength discriminators, the devices employed to encode the incoming polarization state into intensity images (polarization modulators), and the sensor technologies used to register them. We consider the instrumental solutions introduced to perform this kind of measurements at different optical wavelengths and from various observing locations, i.e., ground-based, from the stratosphere or near space.

\end{abstract}

% Include a list of up to six keywords after the abstract

% Include email contact information for corresponding author
{\noindent \footnotesize\textbf{*}F. A. Iglesias,  \linkable{franciscoiglesias@frm.utn.edu.ar} }

\begin{spacing}{2}   % use double spacing for rest of manuscript

\section{Introduction: Instrumental goals and challenges in modern solar spectropolarimetry}
\label{sect:intro}  % \label{} allows reference to this section

Understanding solar activity is important from an astronomical point of view and also from a practical perspective, due to the strong influence the Sun has on Earth and on many human activities carried out both on the ground and in space. Our host star affects Earth's climate \cite{solanki2013, gray2010} and is also the main driver of space weather, which describes the conditions in Earth's magnetosphere and upper atmosphere. Violent solar phenomena can have severe effects on diverse human technology (in particular space-born), like radio communications, various defense and global positioning systems, geostationary and mid-orbit satellites, power distribution grids, railway systems, oil distribution infrastructure, etc.\cite{bothmer2007}. The main driver of solar activity is the highly dynamic solar magnetic field. The most evident manifestation of this is the 11-years sunspot cycle\cite{solanki2003}. At shorter time-scales, hours to minutes, abrupt energetic events such as solar flares or eruptions in the form of coronal mass ejections (CMEs) strongly affect the heliosphere where Earth is embedded\cite{zhang2018}. Moreover, very-small-scale phenomena, in the tens of km range but with global implications on the energetic condition of the different layers of the solar atmosphere, such as magnetic reconnection or wave dissipation, frequently occur both in active and quiet regions of the Sun.

There are regular, in-situ magnetic measurements carried out near Earth, at the Lagrangian point L1, but only few cases 
of this being done closer to the Sun. Two exciting examples of the latter are the recent Parker Solar Probe 
\cite{fox2016,velli2018} and the soon-to-be-launched Solar Orbiter missions \cite{muller2013}. Therefore,  we mostly 
rely on remote sensing to routinely probe the conditions in the different layers of the solar atmosphere\footnote{The 
solar atmosphere is typically divided in four main layers, the lowest layer is the \textit{photosphere} that is few 
hundred km thick and most of the light escapes from the Sun into outer space. The second layer is the 
\textit{chromosphere}, which extends for about 2000 km and presents a slight temperature increase from the photospheric 
5000 K to 7000 K, approximately. The highest layer is the \textit{corona} that has a still-unexplained temperature of 
order 1 MK. In between the corona and chromosphere is the \textit{transition region} where density and temperature 
abruptly change \cite{stix2002}.}.
This implies extracting from the emitted solar radiation, detailed information about temperature, plasma velocity and 
most notably the magnetic field vector, among others. Such information is encoded in the intensity and polarization 
profiles of the Fraunhofer spectral lines, via various radiation--matter--magnetic-field interaction processes 
\footnote{We exclude from our discussion remote-sensing techniques that can derive information about the magnetic field 
using only the observed oscillations of solar features. This is done because they do not rely on polarimetric 
measurements and are limited only to very specific cases where such oscillations are observed \cite{ballester2014}.}. 
Different spectral lines across the solar spectrum are used to probe the various atmospheric layers and may require 
different instrumental solutions and data analysis techniques. We can denote the spectral observational regimes used in 
solar polarimetry as follows:
 
\begin{itemize}
	
	\item \textit{High-energy}: Observations of \textit{linear polarization} at X-ray wavelengths from solar sources have been performed from space since the 70's\cite{tindo1971}. The preferred targets are energetic solar events, most notably flares\cite{zhitnik2006}, where the emissions are related to highly accelerated particles. Instrumental developments during the last decade materialized in successful missions that improved the quality of polarization measurements in terms of polarimetric sensitivity, energy range (e.g., to Gamma-ray\cite{duncan2016}) and spatial resolution, see e.g., Refs. \citenum{mcconnell2002}, \citenum{steslicki2016} and \citenum{shih2012}. High-energy solar polarimetry has many advantages with respect to the other spectral regimes\cite{zharkova2010, kuznetsov2010}. On the other hand, it presents moderate spatial resolution ($\sim$10 arcsec\cite{shih2012}) and polarimetric sensitivity \footnote{Sensitivity in solar polarimetry is used to define the minimum detectable polarimetric signal. In a well calibrated optical instrument, this is directly taken as the root-mean-square (RMS) noise in the Stokes images produced by photon statistics. Polarimetric sensitivity is typically expressed as a fraction of the mean intensity value in the nearest continuum spectral point.} ($\sim 1\%$\cite{duncan2016}), and can not be used to retrieve the complete magnetic field vector.
		
	\item \textit{Microwave and radio}: Mainly \textit{circular polarization} signatures of cm to sub-mm solar radio emissions have been used to retrieve magnetic information about the solar chromosphere and corona. This kind of radio polarimetric measurements can be conveniently done from the ground and have various advantages with respect to the other spectral regimes\cite{casini2017, grebinskij2000}. Notable examples are the detections done with the Nobeyama Radio Telescope\cite{nakajima1985}, e.g., Refs. \citenum{huang2008} and \citenum{miyawaki2016}. On the other hand, radio measurements can be difficult to interpret, they can not be used to retrieve the full magnetic field vector, and have moderate spatial resolution and height information \cite{loukitcheva2017}. In recent years, solar radio magnetometry have gained great impulse due to the increased spatial resolution and sensitivity that can be provided by large interferometric radio telescopes. A prominent example of the latter is the soon-to-be-commissioned mode to observe solar polarization signals with the Atacama Large Millimeter Array (ALMA \cite{wedemeyer2018}). ALMA could reach resolutions in the 0.005 to 5 arcsec range and a polarimetric sensitivity below 0.1$\%$, see Ref. \citenum{loukitcheva2017} for a discussion on the implications for solar polarimetry.
			
	\item \textit{Optical}: Besides the important and complementary radio and high-energy observational regimes, the still dominant technique used nowadays to routinely derive photospheric and chromospheric, high-spatial-resolution, full-vector, magnetic field maps of the Sun, is the inversion of spectropolarimetric data acquired in the optical range of the solar spectrum. The inversion process typically involves the iterative fitting of a predefined atmospheric model to the measured full-Stokes spectral profiles, see the reviews in Refs. \citenum{bellotrubio2006} and \citenum{deltoroiniesta2016}. The most relevant polarizing mechanisms in this regime are the Zeeman and Hanle effects, which can be used to retrieve detailed magnetic field information in various solar conditions, from quiet to active, where magnetic fields range from few G to kG, see e.g., Refs. \citenum{stenflo2013}, \citenum{stenflo2017} and \citenum{deltoroiniesta2003} for some general overviews on the field of solar spectropolarimetry.
		
\end{itemize}

 Given the pros and cons named above for each spectral regime, and considering that the instrumental solutions are substantially different among them, we will devote the rest of this work to optical spectropolarimetry only. The latter plays a central role in modern solar physics and is an important design-driver in any competitive optical solar observatory. For example, four out of the five first-light instruments in the world's largest future solar observatory, the 4m-class Daniel K. Inouye Solar Telescope (DKIST\cite{mcmullin2016}), will have spectropolarimetric (and optional spectroscopic) capabilities \cite{elmore2014,harrington2017}. 

Polarimetry is frequently reviewed using different approaches. Examples of relevant works are given in Ref. \citenum{lagg2015}, that includes a science-driven review on photospheric and chromospheric magnetometry. Ref. \citenum{keller2015} overviews polarimetric instrumentation for the broader field of astronomy. Ref. \citenum{snik2014} summarizes polarization devices and methods used across various disciplines.  In this review we focus on instrumental technology and techniques employed by the solar community to satisfy the demand for data with increasing polarimetric sensitivity and resolution, partially driven by the next generation of large-aperture solar telescopes like DKIST. We will approach the review by quantitatively comparing relevant working and under-construction spectropolarimeters to highlight instrumental concepts that have appeared due to the advances in related fields such as polarization devices manufacturing, imaging sensors technology, data acquisition systems and snapshot spectroscopy, among others. The next sections are devoted to the three main components that form a spectropolarimeter, namely, the wavelength discriminator used to select the desired spectral band; the polarization modulator employed to encode the polarization information typically into temporal, spatial or spectral variations of the output intensity; and the scientific cameras used to detect the modulated intensity signal.

\subsection{Data requirements and trade-offs}
\label{sect:data_req}

\begin{figure}
	\begin{center}
		\begin{tabular}c
			\includegraphics[height=8.5cm]{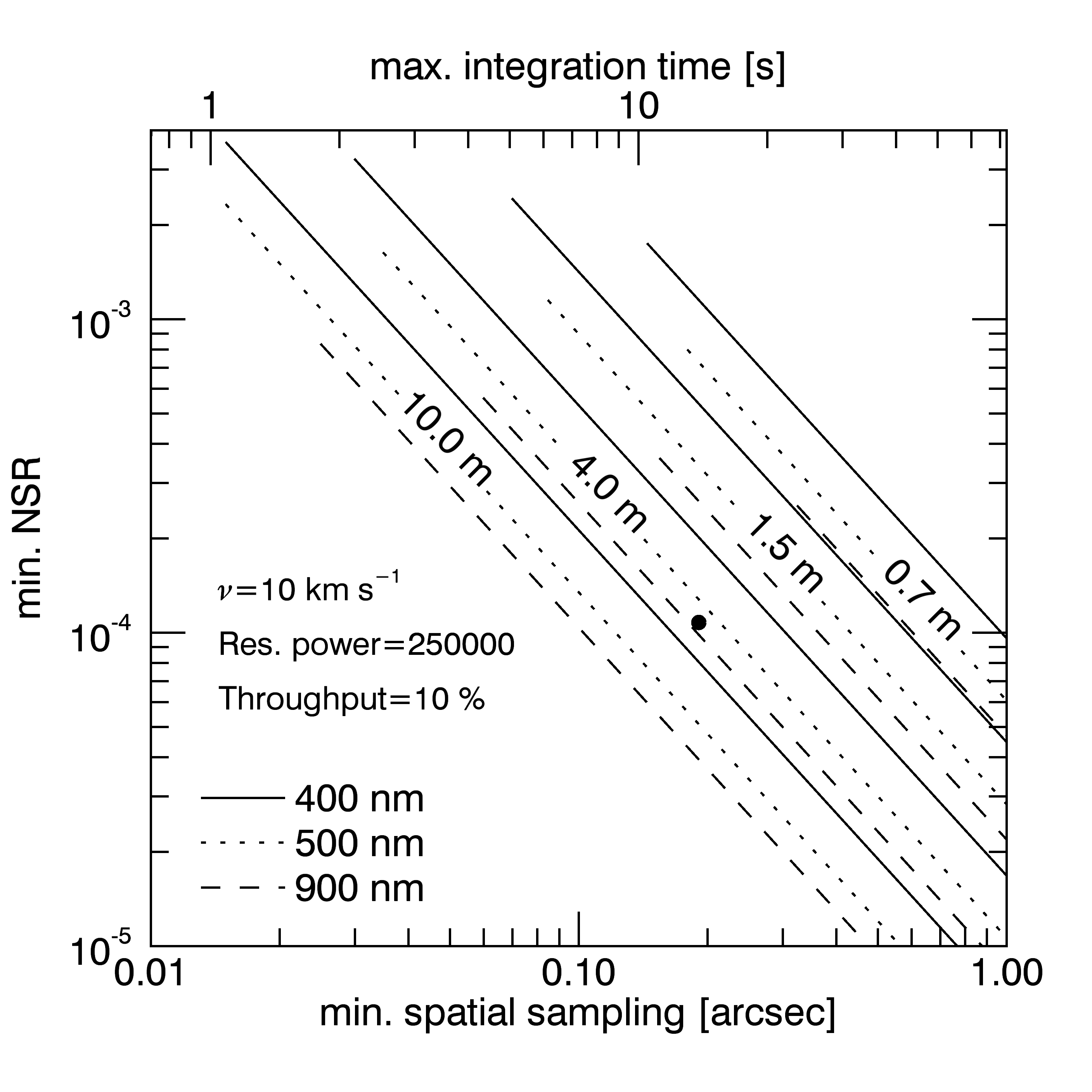}
		\end{tabular}
	\end{center}
	\caption[Trade-off among noise to signal ratio and spatial resolution when doing imaging, full-Stokes spectropolarimetry of a moving solar feature.]
	{ \label{fig:tihrs}
		Trade-off among noise to signal ratio (NSR) and spatial resolution when doing imaging spectropolarimetry of a moving solar feature. Considering the limited flux of the typical solar spectrum and assuming a given wavelength (different line styles, see legend), a fixed aperture (groups of three lines, see annotations) and spatial sampling (lower axis), the maximum integration time (upper axis) is limited, if blurring and polarimetric artifacts due to signal motion are to be avoided. This in turn limits the minimum NSR that can be achieved in the Stokes images (left axis). The assumed optical throughput, spectral resolving power and velocity of the solar feature are given in the legend. We considered ideal polarimetric efficiencies. For a fixed size of the resolution element, the only way to achieve the high-sensitivity ($0.01\%$), high-resolution ($0.2$ arcsec) regime, marked with the black dot, is by increasing the telescope aperture, and thus the photon collecting area, to a 4-m class. However, note that this means working at a resolution much lower than the telescope difraction limit. See the text for extra details. Adapted from Ref. \citenum{iglesias2016b}.} 
\end{figure}

There are different data requirements that arise from the main ongoing research areas in solar polarimetry. The resulting trade offs when designing the required instrumentation which are mainly associated to the high-dimensionality of the data, the required signal-to-noise-ratio (SNR) levels and resolution limitations (see below). Such requirements can be grouped as follows\cite{lagg2015}:
\begin{itemize}
	\item \textit{Simultaneous high polarimetric sensitivity and spatial resolution}: Required to study the ubiquitous, small-scale (tens of km on the Sun\footnote{The density scale height that is a fundamental length for many basic physical processes in the Sun is $\sim 150 km$ at the photosphere.}), faint ($<$10 G) fields present in the photosphere and chromosphere which are critical to understand processes with global energetic implications such as magnetoconvection \cite{borrero2017,jess2016}. This translates in a requirement of sub-arcsec angular resolution, few times $0.01\%$ polarimetric sensitivity\cite{iglesias2016,martinez-pillet2011} and mid-high cadence (see Fig. \ref{fig:tihrs}), which is challenging from an instrumental point of view.
	
	\item \textit{High polarimetric sensitivity of faint signals}: In this case, spatial and spectral resolution need to be reduced to reach the required sensitivity for studying weak magnetic fields, e.g., Refs.  \citenum{lites2004,beck2005c,lagg2007,lagg2016}, particularly in the corona where the effective photon flux is low. Tomographic inversions or line-ratio based methods can be used to quantify coronal fields from polarization measurements acquired with coronagraphs\footnote{A coronagraph is an instrument that observes the corona by simulating a total solar eclipse using an artificial occulter in front or inside the telescope. Since the corona is very dim with respect to the solar disk, by a factor of $\sim 10^{-7}$ in white light, stray-light rejection is the driving design criterion \cite{rougeot2017}. Coronagraphs are built with polarimetric capabilities to e.g., study the so-called K-corona that is globally linearly polarized in the $10\%$ level due to the radiation anisotropy.} or via off-limb observations. Coronagraph measurements have been done in the infrared (IR) \cite{lin2004, dalmasse2016, kramar2016} and will be tried soon for the ultraviolet (UV)\cite{pagano2015}.  Examples of off-limb magnetometry in the near IR (NIR) line at 1083.0 nm are given in Refs. \citenum{schad2016}, \citenum{schad2018} and \citenum{kramar2016} which employs a vector tomographic reconstruction. Due to the difficulties to obtain coronal spectropolarimetric observations, these fields are also frequently guessed via extrapolations of photospheric data (see below) or constrained magnetohydrodynamic (MHD) simulations.
	
	\item \textit{Full-disk spectropolarimetry}: In this case the field of view (FOV) covers the full solar disk with moderate spatial resolution ($\sim 1$ arcsec) and cadence (minutes). Traditionally, only circular polarization was measured to obtain maps of the line-of-sight (LOS) component of the magnetic field, e.g., as done in the notable Michelson Doppler Imager (MDI \cite{scherrer1995}) on board the 20-years-running Solar and Heliospheric Observatory (SOHO\cite{domingo1995}). However, nowadays the preference is to record the full Stokes vector, e.g., as done in the successful HMI listed in Table \ref{tab:all_pol}. The resulting synoptic maps of plasma velocity and vector magnetic field, among others, are widely used in space weather programs to e.g., derive the coronal magnetic field from photospheric extrapolations \cite{wiegelmann2017}. To provide continuous coverage from the ground, typically a network of identical instruments located around the globe is employed\cite{balasu2011}. Two notable examples are the Oscillations Network Group (GONG \cite{harvey1996}), that only measures circular polarization, and the upcoming European-funded, Solar Physics Research Integrated Network Group (SPRING\cite{roth2017}) which will provide the full Stokes vector.   
	
	\item \textit{High-cadence spectropolarimetry}: The study of fast solar events such as flares \cite{kucklein2015, kuridze2018, kleint2017b} and CMEs or filaments eruptions \cite{deforest2017,wang2015}, requires sometimes measurements of the full Stokes vector with a cadence of few minutes or seconds.
\end{itemize}

Performing the above-described, spectropolarimetric measurements requires estimating the direction, energy, 
time-of-arrival and polarization of the incoming photons. This derives in a measurement space with five dimensions 
(plane-of-sky coordinates $x$ and $y$ ,wavelength $\lambda$, time $t$ and Stokes vector $S=$[$I$, $Q$, $U$, $V$]) that 
has to be mapped to a three-dimensional data space that represents the acquisition done with a two-dimensional detector 
at a given instant of time ($x_d$, $y_d$ and $t_d$). Since the data space is of lower dimensionality, it is necessary 
to encode more than one measurement dimension in a single data parameter. This implies a trade-off in terms of 
simultaneity, resolution and/or FOV among the measurement dimensions that are sharing a single data parameter. The 
existing solutions, summarized in Table \ref{tab:sp_map}, generally divide this process in two steps corresponding to 
the spectroscopic and polarimetric analyses, see Sec. \ref{sect:mod} and \ref{sect:spect} respectively. Recent 
instrumental concepts have been developed to perform the complete spectropolarimetric analysis in a single device, 
although with limited performance (see Sec \ref{sect:sp_mod}).

\begin{table}[!htp]
	\caption[Summary of possible spectropolarimetric mappings.]{Summary of possible spectropolarimetric mappings. Selected combinations when trying to map the five-dimensional, spectropolarimetric measurement space given in column one, into the three-dimensional data space provided by the scientific detector, ($x_d,y_d,t_d$), are given in columns 4 to 6. We have put in parentheses the measurement dimensions that, due to optical constraints, must be imaged in the same detector area, e.g., when doing spatial polarization modulation of a spectrograph output $\lambda$ and $y$ must be imaged in the same detector area while the different $S$ could be imaged in a separate camera. Extra information about the spectral and polarimetric parts of the mapping are given in Sec. \ref{sect:spect} and \ref{sect:mod} respectively. In general, the more a given data parameter is populated by different measurement dimensions, the stronger are the trade-offs in terms of simultaneity, resolution and/or field of view. We have highlighted in bold the combinations that allow snapshot-spectropolarimetry of extended sources which, among other benefits, maximizes SNR by making use of all the relevant photons reaching the instrument at a given time.}
	\centering
	\small
	\begin{tabular}{|c | l l | c c c |}
    \noalign{\smallskip}     		
    \cline{1-6}  
	%\noalign{\smallskip}  
	Meas. Space & \multicolumn{2}{c|}{Spectropolarimetric Mapping} & \multicolumn{3}{c|}{Data Space} \\
	\cline{1-6}  
	 & Spectral & Polarimetric &$x_d$ &$y_d$ &$t_d$ \\ 
	\cline{2-6} 
%	\noalign{\smallskip}  	
	 & Filtergraph & Temporal &$(x)$ &$(y)$ &$t,\lambda,S$ \\ 
	 & 			   & Spatial  &$(x),S$ &$(y),S$ &$t,\lambda$ \\ 
	 & 			   & Spatio-temporal  &$(x),S$ &$(y),S$ &$t,\lambda,S$ \\ 
	\cline{2-6}
%	\noalign{\smallskip}  	
	 & Spectrograph & Temporal &$(\lambda)$ &$(y)$ &$t,x,S$  \\ 
 	 $x,y,\lambda,t,S$& 			   & Spatial   &$(\lambda),S$ &$(y),S$ &$t,x$  \\ 
 	 & 			   & Spatio-temporal   &$(\lambda),S$ &$(y),S$ &$t,x,S$  \\ 
	\cline{2-6}	     
%	\noalign{\smallskip}  
	 & Integral Field & Temporal &$(x,y,\lambda)$ &$(x,y,\lambda)$ &$t,S$  \\
	 & 			      & \textbf{Spatial}  &$(x,y,\lambda),S$ &$(x,y,\lambda),S$ &$t$  \\ 
	 & 			      & Spatio-temporal &$(x,y,\lambda),S$ &$(x,y,\lambda),S$ &$t,S$ \\ 
	\cline{2-6}	 
%	\noalign{\smallskip}  	
	 &  \multicolumn{2}{c|}{\textbf{Spectropolarimetric modulation$*$\tnote{*}}} &$x,y,\lambda,S$ &$x,y,\lambda,S$ &$t$  \\ 
    \cline{1-6}		
\end{tabular}
 \begin{tablenotes}
	\footnotesize
	\item[*] * In this case, which groups several techniques, a function (e.g., fringes pattern) of the measurement parameters is commonly mapped to the detector.
\end{tablenotes}
\label{tab:sp_map}
\end{table}

Besides the trade-offs that arise from the high dimensionality of the measurement space shown in Table 
\ref{tab:sp_map}, there is an intrinsic limitation to the achievable SNR when imaging any moving solar signal at high 
resolution. Because of the limited solar flux and well capacity of imaging sensors, the accumulation of many frames is 
required to increase polarimetric sensitivity. E.g., to reach 1 G of longitudinal magnetic sensitivity approximately 
$10^7$ photons are required\cite{martinez-pillet2011}. The latter, in turn, may result in blurring and polarimetric 
artifacts due to the signal movement from one sampling element to the adjacent one during the integration time, thus 
effectively reducing the final spatial and spectral resolutions. This trade-off is exemplified in Fig. \ref{fig:tihrs}, 
where we show the minimum combined noise to signal ratio (NSR) and spatial resolution that can be reached when doing 
full-Stokes, imaging spectropolarimetry of a solar feature that is moving close to the photospheric sound speed ($\sim 
10$ $km s^{-1}$) with a given telescope aperture and wavelength. We have used an average solar spectrum, a conservative 
$10\%$ total optical throughput of the system, ideal polarimetric efficiencies\footnote{Polarimetric efficiencies 
quantify the noise propagation in the polarimeter given its adopted modulation scheme, see e.g., Refs. 
\citenum{deltoroiniesta2000, deltoroiniesta2003} and \citenum{deltoroiniesta2012}.} and a competitive spectral 
resolution of 250000. Note that the most demanding observing regimes can only be achieved by increasing the telescope 
photon collecting area (aperture) while keeping the sampling element size well below the one defined by the diffraction 
limit, i.e., the spatial resolving power must be sacrificed to increase SNR.

\subsection{State-of-the-art and upcoming solar optical spectropolarimeters}
\label{sect:opt_pol}

\begin{figure}	
	\begin{center}
		\begin{tabular}c
			\includegraphics[height=8.5 cm]{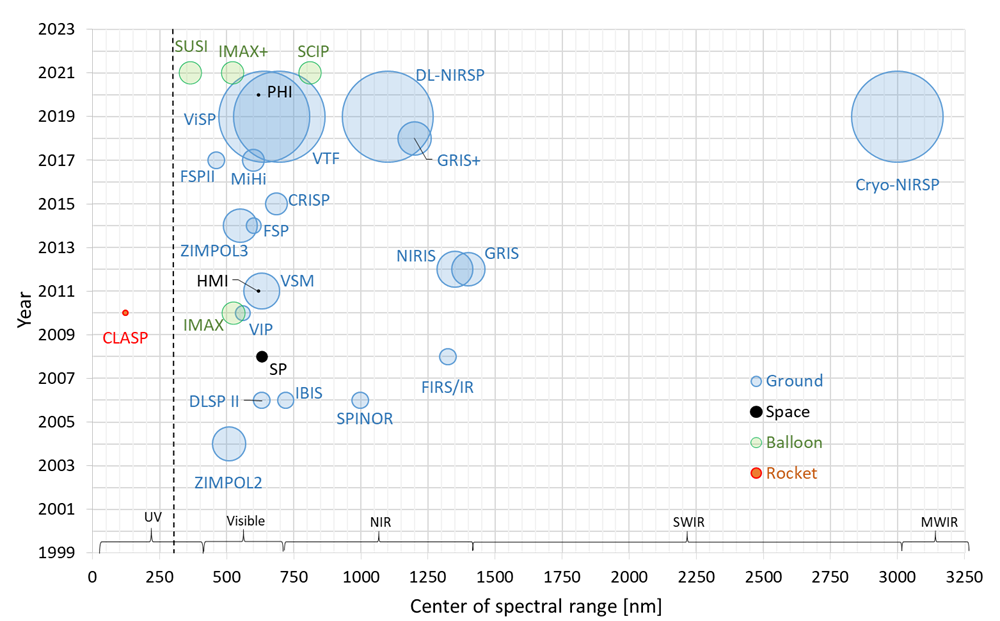}
		\end{tabular}
	\end{center}
	\caption[Date of introduction, aperture and spectral regime of the instruments listed in Table \ref{tab:all_pol}.]{Date of introduction (vertical axis), aperture (proportional to the bubbles radii) and center of the spectral range (horizontal axis) of the instruments listed in Table \ref{tab:all_pol}. The largest aperture corresponds to the first-light instruments of the upcoming 4-m DKIST\cite{elmore2014}. The different observatory locations are highlighted using colors, see legend. The vertical dashed line denotes the atmospheric cutoff at $\sim$310 nm}
	\label{fig:all_pol_aperture}    
\end{figure}

In this section we list selected and representative, solar optical spectropolarimeters that were developed in the last 
decades or are currently under construction. These are presented in Table \ref{tab:all_pol} along with references. 
Additionally, Fig. \ref{fig:all_pol_aperture} illustrates basic properties such as the approximate year of 
introduction; aperture and location of the telescope; and center of the spectral coverage. Other relevant properties 
will be presented and discussed in the following sections. The specifications of all the polarimeters dated 2018 or 
beyond were taken from the reported design parameters. Note that the extracted properties are meant to give an idea of 
the instrumental overall capabilities and cannot capture the large range of specifications that in some cases is 
provided by the configuration flexibility of these instruments (the corresponding references can be consulted for 
further details). We also note that we have excluded from this review some successful earlier instruments, most notably 
the Advanced Stokes Polarimeter (ASP\cite{elmore1992,skumanich1997}) and the Tenerife Infrared Polarimeter 
(TIP\cite{pillet1999, collados2007}), as well as the THEMIS\cite{ariste2000} telescope. All have made substantial 
contributions to the field and are well reported elsewhere. ASP and the second version of TIP are not operational 
anymore, their technology have been the base for the design of other relevant instruments such as DLSP or Hinode SP for 
the former and GRIS for the later, see Table \ref{tab:all_pol}. THEMIS, which presents a unique, polarization-optimized 
design, does not have a polarimeter listed among the available instrumentation in its 2019 observing run.

Many of the listed instruments work in the visible range of the spectrum, where high-resolution polarimetry was firstly 
developed mainly because it can be accessed from the ground, it presents high flux and simplifies the optical design 
with respect to the UV and IR regimes. The visible range also contains many photospheric lines which are used to probe 
the strongest solar magnetic fields, e.g., in sunspots, and are normally easier to interpret than chromospheric lines. 
Observations in the NIR and short-wavelength IR (SWIR) regimes are useful because the  the Zeeman splitting (in units 
of Doppler broadening) increases linearly with wavelength. On the other hand, the spatial resolution in the IR is 
reduced and the instrument design is more complex due to the worse response of detectors, which typically needed to be 
Nitrogen cooled (see Sect. \ref{sect:detect}), and optical components. The UV and near UV (NUV) windows are also poorly 
explored mainly because they are partially not accessible from the ground (see below) and the lower solar flux and 
shorter wavelengths impose strict constraints to the optical components and detectors in terms of photon efficiency, 
wavefront distortions and straylight\cite{gandorfer2006}. In addition, the available polarization modulation solutions 
in the UV are more limited (see e.g., Ref. \citenum{larruquert2017}). However, spectral lines in the UV are in general 
much steeper than at longer wavelengths, due to the steepness of the Planck function in the UV. This increases the 
Zeeman signals and can counteract the effect of the reduced Zeeman splitting to a significant 
degree\cite{riethmuller2019}. Observations below the atmospheric cutoff ($\sim 310$ nm) cannot be done from the ground. 
Moreover, all the space-born, solar polarimetric measurements that have been acquired to date were done in the visible. 
The only imaging spectropolarimetric exploration of the UV regime was done by the 2015 flight\cite{kubo2017} of CLASP, 
see Table \ref{tab:all_pol}. CLASP is a spectrograph-based, UV polarimeter fed by a 27.9 cm telescope mounted in a 
sounding-rocket. It can observe linear polarization (Stokes I, Q and U) around 121.1 nm to study scattering 
polarization for about 5 min during each flight and its development, led by the Japanese Space Agency, was meant mainly 
as a demonstrator for a future space mission.  The poor polarimetric knowledge of the UV and NUV windows is one of the 
main motivations for the development of SUSI, see Table \ref{tab:all_pol}, to be included in the third flight of the 
ballon-borne SUNRISE observatory\cite{solanki2017,barthol2018}.

As can be appreciated in Fig. \ref{fig:all_pol_aperture}, the available solar observatories have up to date apertures below 2 m. Moreover, the technical challenges and large costs involved, have limited the aperture size of space-born telescopes to a fraction of this, namely 0.5 m for the HINODE/SOT. The latter, along with the upcoming 4-m DKIST (first light expected in 2019) and the important improvements made in multi-conjugated solar adaptive optics systems \cite{zhang2017,stangalini2018,schmidt2018,rao2018} and image restoration techniques\cite{lofdahl2007,sutterlin2000, vannoort2017,denker2018,suzuki2018,asensioramos2018}, have slightly shifted the scale in favor of ground-based observatories when trying to reach simultaneously the highest spatial resolution and SNR\footnote{The SOLAR-C mission, equipped with a powerful 1.4-m telescope and outstanding spectropolarimetric capabilities, was envisioned as a natural successor of the successful HINODE by the Japanese Space Agency, and proposed by an international consortium to its European counterpart in 2015\cite{solanki2015b}. Even though the proposal got high marks, it was not selected and funding is not clear as of today\cite{kleint2017}.}. As a consequence, two advantages of space solutions become relevant, namely the ability to observe wavelengths absorbed by Earth's atmosphere and to monitor the Sun continuously with diffraction limited performance (crucial to answer many science questions).

\begin{table}[!htp]
	\caption [Selected working and upcoming solar optical spectropolarimeters analyzed are in this review.]{Selected working and upcoming solar optical spectropolarimeters analyzed in this review along with their approximate working wavelength range or points. Note that some instruments may cover non-listed spectral ranges using different configurations, e.g., cameras, or not cover the full range, e.g., filtergraphs which typically rely on pre-filters availability. When relevant, we have used only the specifications of the modes that can measure the full Stokes vector. The instruments marked with an asterisk are polarimeters only and in general can be operated with different telescopes and/or wavelength discriminators. In this review we use the specifications of the referenced implementations.}
	\centering
	\small
	\begin{tabular}{l l l l l}
		\noalign{\smallskip}     		
		\cline{1-5}  
		\noalign{\smallskip}  
		\# & Observatory&Instrument & Wavelength range [nm] & Reference \\
    	\noalign{\smallskip}  
		\cline{1-5}  
		\noalign{\smallskip}  
		1&McMath-Pierce&ZIMPOL2* & 316-700 & \citenum{gandorfer2004} \\
		2&DST&IBIS & 580-860 & \citenum{cavallini2006,viticchie2009,viticchie2010}\\
		3&DST&DLSP II & 630.25 & \citenum{sanka2004,sanka2006}\\
		4&DST&SPINOR & 430-1565 & \citenum{socnav2006,beck2019}\\
		5&DST&FIRS-IR & 1083; 1565& \citenum{jaeggli2008,jaeggli2010}\\
		6&DST&FSPII* & 450-750 & \citenum{zeuner2019}\\			
		7&BBSO-NST&VSM & 630.2& \citenum{keller2003, vsm_website, balasu2011}\\
		8&BBSO-NST&NIRIS & 1000-1700& \citenum{cao2012}\\
		9&SST&CRISP& 510-860 & \citenum{scharmer2008,delacruzrodriguez2015}\\
		10&SST&MiHi &450-750 & \citenum{mihi_website,vannoort_PC2018,jurcack2018}\\		
		11&VTT&FSP* &400-800& \citenum{iglesias2016, feller2014}\\
		12&VTT&VIP & 420-700& \citenum{beck2010}\\		
		13&GREGOR&ZIMPOL3* &400-700 & \citenum{ramelli2010,ramelli2014}\\	
		14&GREGOR&GRIS& 1000-1800 & \citenum{collados2012,schmidt2012}\\
		15&GREGOR&GRIS+* & 800-1600 & \citenum{doerr_PC2018} \\					
		16&DKIST&ViSP &380-900& \citenum{dewijn2012,nelson2010,dkist_website}\\
		17&DKIST&VTF &520-870& \citenum{schmidt2014c,kentischer2012,schmidt2016,dkist_website}\\
		18&DKIST&DL-NIRSP & 500-1700 & \citenum{jaeggli2017,elmore2014,dkist_website}\\
		19&DKIST&Cryo-NIRSP &1000-5000 & \citenum{elmore2014,dkist_website}\\
		20&SUNRISE 1\&2&IMAX & 525.02& \citenum{martinez-pillet2011, barthol2011}\\		
		21&SUNRISE 3&IMAX+ & 517.3; 524.70; 525.02& \citenum{solanki2016_sunrise3, barthol2018, imaxplus_specs}\\
		22&SUNRISE 3&SUSI &300-430 & \citenum{solanki2016_sunrise3, barthol2018}\\
		23&SUNRISE 3&SCIP& 765;855 & \citenum{solanki2016_sunrise3,barthol2018, suematsu2018}\\
		24&HINODE-SOT&SP & 630.15; 630.25& \citenum{lites2013,ichimoto2008 }\\	
		25&SDO&HMI & 617.3 & \citenum{wachter2012,scherrer2012,schou2012,couvidat2012}\\				
		26&CLASP& CLASP &121.6& \citenum{kobayashi2010,kubo2017}\\
		27&SO&PHI& 617.3 & \citenum{gandorfer2011,solanki2015}\\		
		\noalign{\smallskip}  
		\cline{1-5}		
	\end{tabular}
\label{tab:all_pol}
\end{table}

\section{Wavelength discriminators}
\label{sect:spect}

\begin{landscape}
\begin{figure}	
	\begin{center}
		\begin{tabular}c
			\includegraphics[height=14 cm]{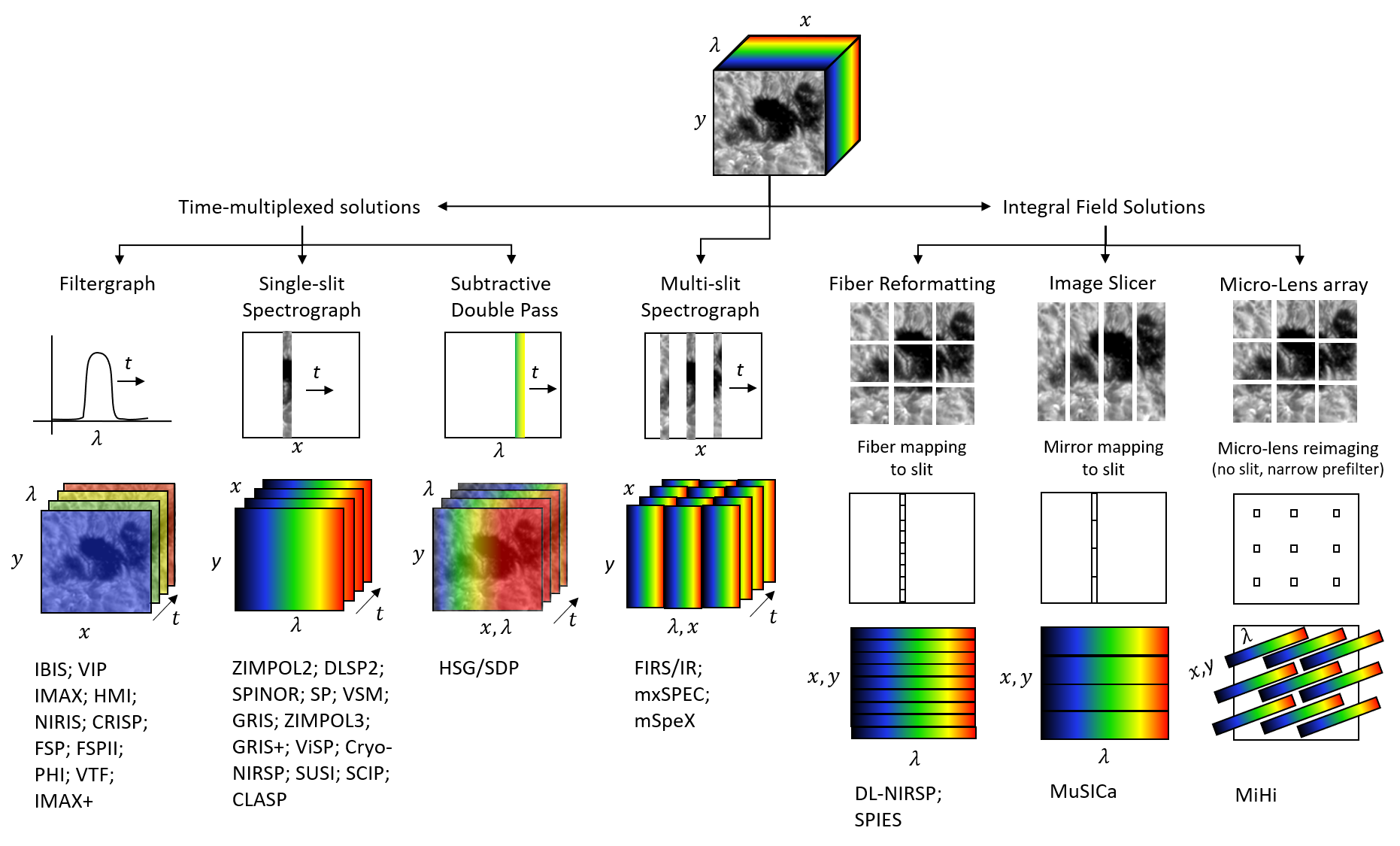}
		\end{tabular}
	\end{center}
	\caption[Seven different spectral mapping techniques used in solar spectropolarimetry]{Six different spectral mapping techniques used in solar spectropolarimetry. The cube ($x$,$y$,$\lambda$) at the top can be mapped to the two-dimensional detector (bottom row) either by time-multiplexing, as done in filtergraphs and spectrographs, or by making use of a larger detector area to accommodate all the measurement dimensions in a single frame exposure, see Table \ref{tab:sp_map}. The latter is used in the integral field solutions, which are mostly under development and can perform snapshot-spectroscopy at the expenses of a reduced resolution and/or FOV. The multi-slit spectrograph can be considered an in-between solution. We have listed example instruments using each technology at the bottom, see Table \ref{tab:all_pol} and the text for extra details.}
	\label{fig:sp_map}    
\end{figure} 
\end{landscape}

\begin{figure}	
	\begin{center}
		\begin{tabular}c
			\includegraphics[height=7 cm]{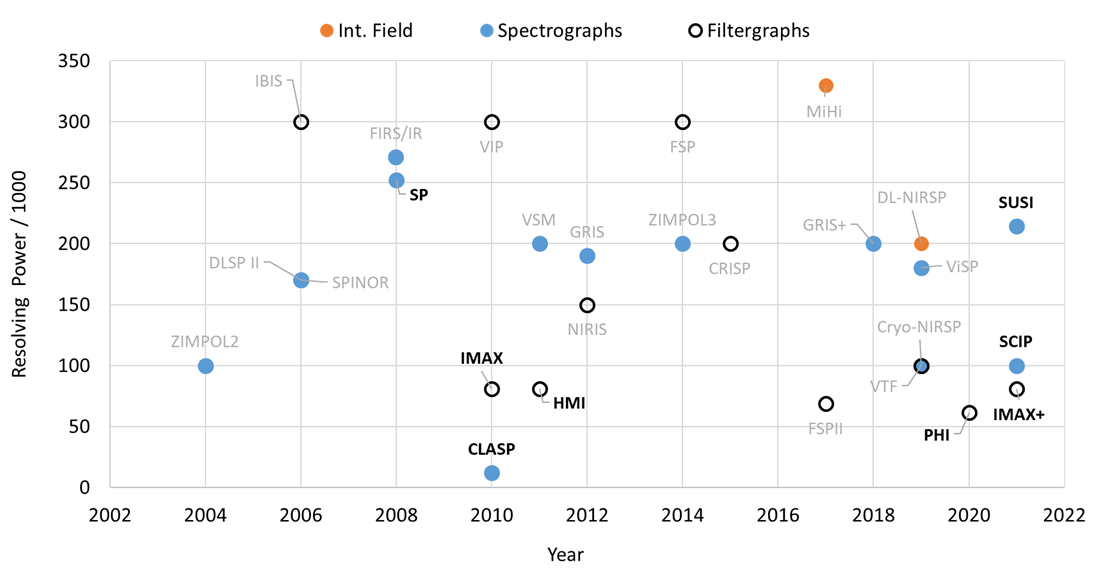}
		\end{tabular}
	\end{center}
	\caption[Spectral resolving power versus date for the instruments in Table \ref{tab:all_pol}.]{Approximate spectral resolving power (vertical axis) versus date (horizontal axis) for the instruments listed in Table \ref{tab:all_pol}. Each labeled dot represents a polarimeter, we use black labels for space, rocket and balloon borne instruments. The different spectral mapping techniques are highlighted using colors, see legend. Note that, the reported resolving power may be the maximum achievable in a given instrumental configuration and/or a fraction of the working spectral range. }
	\label{fig:sp_res}    
\end{figure}

The wavelength discriminators that have been most successfully used to perform the spectral mapping (see Table \ref{tab:sp_map}) in optical solar spectropolarimetry are grating spectrographs (SGs) and filtergraphs (FGs) systems, see Fig. \ref{fig:sp_map}. Historically, the instrumental developments focused on minimizing optical aberrations and maximizing throughput, accuracy and spectral resolution. As a consequence, nowadays the richness of typical photospheric and chromospheric spectral line profiles can be properly sampled using both approaches, e.g., 21 of the 27 instruments in Table \ref{tab:all_pol} have a spectral resolving power above 100000, with 13 using SGs and 11 FGs, see Fig. \ref{fig:sp_res}. Recent designs focus on exploiting the ability to observe multiple lines simultaneously, reducing scanning times, polarimetrically exploring poorly known portions of the spectrum and/or increasing spatial resolution. Additionally, the inability of SGs and FGs to simultaneously capture both spatial and spectral information is being tackled by developing integral field solutions. More details on each of the approaches are given in the following subsections.

\subsection{Spectrographs}
\label{sect:sg}

Given the high spectral resolution required, SG-based solar polarimeters employ almost exclusively echelle gratings to 
maximize throughput. The preferred solution to effectively reduce one spatial dimension at the SG input is using a long 
slit. The slit substrate can also be used to reflect the light, corresponding to the unused portion of the FOV, to feed 
a context wide-band, slit-jaw imager\footnote{Or, ideally, a parallel FG-based polarimeter, although this has not been 
tried \cite{lagg2015}}. Slit SGs capture the full spectrum of the one-dimensional slit in a single exposure. However, 
they have to scan the solar surface to generate a two-dimensional map of an extended source. As a consequence, the 
resulting spatial information is not simultaneous in the scanning direction, and its quality is generally affected by 
the slit width and accuracy of the scanning system. Additionally, the required scanning time can easily be larger than 
the solar evolution time, when aiming for high sensitivity and spatial resolution measurements of large targets such as 
sunspots.However, SGs have been employed for synoptic observations (e.g., VSM), where SNR and spatial resolution 
requirements are less strict.

Due to the above-mentioned properties, SGs are usually considered as the low-spatial-resolution option when compared to FGs. Such a difference is even stronger in ground-based systems, were numerical image restoration techniques have been developed to reduce atmospheric seeing effects, and substantially improve the final spatial resolution of FG data, see e.g., the review in Ref. \citenum{lofdahl2007}. Image restoration of ground-based SG data using traditional techniques has proven difficult, mainly because of the longer integration times involved and the absence of full spatial information.  Previous works, see Refs. \citenum{keller1995, sutterlin2000} and \citenum{beck2011}, paved the way to a very recent development, see Ref. \citenum{vannoort2017}, that overcomes these limitations and uses the simultaneously recorded slit-jaw images, to estimate post-facto the seeing degradations and restore the spectra via a multi-frame blind deconvolution\cite{vannoort2005}. This restoration allows reaching a resolution close to the optical diffraction limit in both spatial dimensions, while preserving the full spectral information and increasing the polarization signal levels. Moreover, it has been applied to the visible spectral range and is currently being tested at SWIR wavelengths with GRIS+, see Table \ref{tab:all_pol}. Important improvements have also been made in multi-line inversion techniques, which allow obtaining better estimates of atmospheric parameters, including sometime their height dependence, by simultaneously fitting the polarimetric profiles of many spectral lines\cite{bellotruibio2005, beck2008, casini2009,beck2011,demidov2012,balthasar2012}. For a given polarimetric sensitivity, making use of multiple lines improves the SNR of the latter inferred parameters and is crucial to reach the high-resolution, high-sensitivity regime named in Sect. \ref{sect:intro} under photon-starved conditions. Due to the latter and other multi-line techniques, such as the line-ratio method\cite{stenflo2013b,khomenko2007}, the simultaneous observation of many lines, sometimes belonging to different spectral windows, has been exploited in the last decades of solar SG developments, see e.g., Refs. \citenum{bellotruibio2005, collados2007,socnav2006,beck2007,beck2011b} and \citenum{lites2013}. Newer designs seek to follow or improve these capabilities in combination with higher spatial resolutions, e.g., ViSP and SPINOR are designed to simultaneously observe up to three and four visible wavelength ranges respectively, and the exploration of the NUV regime aimed by SUSI will include a many-lines inversion, see Ref. \citenum{riethmuller2019}.

\subsection{Filtergraphs}
\label{sect:fg}

Different devices can be used to obtain narrow-band filtergrams, the ones that have been most successfully employed in solar spectropolarimetry are Michelson and Fabry-Perot interferometers (FPI's). The former were  selected for the successful space-born polarimeters MDI and HMI, see Table \ref{tab:all_pol}, mainly due to their better stability and smaller size\cite{scherrer1995}. However, Michelson interferometers are not suitable to observe multiple spectral lines. FPI's are highly reflective cavities with a known, tunable thickness. The most used types in ground-based devices are air-gap etalons which can be rapidly tuned using piezo-elastic actuators\cite{prasad2002}. For balloon or space applications, crystal-based FPI's such as those based on solid lithium niobate wafers are preferred because they are lighter, more stable and their resulting apertures smaller\cite{martinez-pillet2011}. FPI's present high transmissions and a broader usable spectral range compared to Michelsons. On the other hand, they require higher voltages to operate and more effort has to be put in their thermal stabilization and tuning control, to ensure a stable response that is also homogeneous across the FOV. In spite of this, single or multiple FPI's are the preferred option in ground-based FGs, and they were the solution adopted by the design team of both the IMaX\cite{martinez-pillet2011} and PHI instruments\cite{gandorfer2011}. Moreover, 10 of the 11 FGs in Fig. \ref{fig:sp_res} use FPI's with the most common configuration (7 instruments) being two FPI's in tandem (or two beam passes of a single device) to increase spectral resolution and throughput (because broader prefilters can be used\cite{beck2010}). A  triple-FPI instrument, the Telecentric Etalon SOlar Spectrometer (TESOS\cite{tritschler2002}) has also been successfully used in combination with different polarimeters, e.g., FSP and VIP, to perform high-resolution solar observations.

FGs record both spatial dimensions simultaneously with similar optical performance, and have to scan in wavelength to sample the spectral line profile, see Table \ref{tab:sp_map}. Due to the former, combining FGs with image restoration techniques and adaptive optics systems is the preferred option to do high-spatial resolution polarimetry of extended solar targets, particularly for fast-changing events like flares. Such a combination presents calibration challenges, particularly for ground-based systems, because different instrumental and ambient effects, e.g., telescope polarization and seeing, can be non-linearly entangled\cite{schnerr2011}.

\subsection{Integral field solutions}
\label{sect:ifs}

During any given frame exposure, both SGs and FGs miss (do not detect) a considerable number of photons that belong to the aimed solar signal, because of their inability to cover the desired spatial and spectral FOVs simultaneously. Such a situation is not desired in high-resolution observations due to the already strict trade-off imposed by the limited flux and solar evolution (see Fig. \ref{fig:tihrs}). This is, along with the reduction of signal smearing and polarimetric artifacts, the main argument to develop integral field techniques for solar spectropolarimetry. Many devices have been developed that are capable of doing snapshot spectroscopy, i.e., mapping the full measurement cube ($x$,$y$,$\lambda$) on the two detector dimensions ($x_d$,$y_d$) in a single exposure, see Table \ref{tab:sp_map} and the review in Ref. \citenum{hagen2013}. This improvement comes at the expenses of a reduced spatial/spectral FOV, along with a more complex optical setup and data reduction. Spatial resolution is typically not sacrificed, for reasons of image restoration which relies on critical sampling. Also spectral resolution is typically not compromised because Stokes inversions rely on a given minimum sampling of spectral lines. Integral field solutions have been applied to night-time astronomical observations since their introduction in the 60's\cite{hagen2013}. On the other hand, the solar community started applying these methods more recently, partially motivated by the availability of sizable imaging detectors that can be used in large-aperture solar telescopes. The following five techniques, sketched in Fig. \ref{fig:sp_map}, are currently being developed for solar spectropolarimetry.

\begin{itemize}

	\item \textit{Multi-slit spectrograph}: The two-dimensional FOV is partially covered using $N$ slits that simultaneously feed a single SG. In order to accommodate the $N$ output spectra in a single detector avoiding any overlap, the slits have to be properly placed and special narrow-band and order-sorting filters need to be employed. Note that this method is not strictly speaking an integral field solution because it does require substantial spatial scanning to image an extended source. However, the scanning time is reduced and the system throughput increased, both by a factor of $N$ compared to a single slit SG. This concept was first tried in solar astronomy in the 70's without polarimetry\cite{martin1974} and further applied in few other instruments, e.g., Ref. \citenum{srivastava1999}. Recent implementations are the FIRS spectropolarimeter (see Table \ref{tab:all_pol}), which can use up to four slits and has two parallel spectral channels to observe simultaneously the lines near 6302 \textit{\AA} in the visible and 10830 or 15648 \textit{\AA} in the IR \cite{jaeggli2010}; the experimental Massively Multiplexed Spectrograph (mxSPEC\cite{lin2014}) that is equipped with a 40-slits, full-disk spectrograph; and the higher-resolution version of the later, mSpeX\cite{schad2018}.

	\item \textit{Subtractive double pass}: Another technique that improves the single SG performance in terms on 2D-spectroscopic capabilities is the Subtractive Double Pass (SDP)\cite{mein1972}. SDP was successfully employed to do solar spectroscopy in the 60's\cite{mein1972, stenflo1986} before the FPI's era. The technique was subsequently improved and has been continuously used up to date, see e.g., Refs. \citenum{mein1991,mein2009,schmeider2004,lopezariste2011,heinzel2015}, although with much reduced frequency than FPIs or traditional slit SG's and has not been considered in newer instrumental designs. In SDP the slit at the SG entrance focal plane is removed and the desired 2D FOV is dispersed by a grating. After passing a slit in the output focal plane of the SG, the beam path is reversed and projected back via the grating into an entrance focal plane. The resulting 2D image shows a continuous variation of wavelength in the dimension perpendicular to the slit, see Fig. \ref{fig:sp_map}. The spectral bandpass is determined by the slit width. By adding the possibility to move the spectral slit, the 3D spectroscopic data cube can be sampled. Partially motivated by the technical difficulties and large price tags involved when developing FPI's systems for large  aperture ($>$1 m) telescopes\footnote{For example, the etalons being developed for the  VTF instrument at DKIST, which are among the largest in the world (250 mm clear aperture with 3 nm rms surface flatness), were very challenging (requiring an industrial consortium) and expensive (several Millions Dollars) to manufacture.}, SDP was recently implemented at the DST telescope by modifying the existing Horizontal Spectrograph (HSG\cite{dunn1991}) to observe in the H$\alpha$ line\cite{beck2018}. Such a demonstration further supports SDP systems as a viable alternative to FPI's when aiming for high-spatial with moderate spectral resolution across a FOV compatible with large-aperture telescope requirements. Note that SDP is a time-multiplexed solution and thus presents the limitations named in Sect. \ref{sect:ifs}.
	
	\item \textit{Fiber reformatting}: This technique densely samples the focal plane, using an integral field unit formed by a bundle of optical fibers. The fibers are reformatted in the exit plane to form one or multiple long slits that are fed to a SG. Many challenges need to be faced to successfully manufacture the fiber bundle while avoiding defect fibers, cross-talk and obtaining a light-efficient coupling. Additionally, standard stock multi-mode fibers do not typically preserve polarization\cite{schad2014}. This technique was successfully implemented in the SpectroPolarimetric Imager for the Energetic Sun (SPIES\cite{lin2012,spies_website}), a testing platform for the upcoming DL-NIRSP (see Table \ref{tab:all_pol}), using a fiber bundle (named BiFOIS-4K\cite{schad2014}) that reformats 15360 cores to four slits. After the SG, the data cube is sampled with 60$\times$64 spatial and 250 spectral pixels using a plate scale of 0.03 arcsec and 43 m\textit{\AA} per pixel respectively. A 90$\%$ yield was obtained with the engineering prototype of BiFOSI-4K, which is $\sim$20$\%$ smaller than the one planned for DL-NIRSP. The latter will also observe in three simultaneous spectral channels, covering 500 – 900 nm, 900–1500 nm, and 1500 – 2500 nm approximately. Due to the difficulty of manufacturing large fiber bundles, a disadvantage of this approach is the limited FOV obtained when using a high-spatial resolution. This is why DL-NIRSP will employ a mirror-based scanning system to cover large objects, like a sunspot, or the full FOV of the 4-m DKIST telescope\cite{elmore2014}.

	\item \textit{Image slicer}: In this case, the focal plane reformatting is done using a stack of thin mirror slices. Each slice reflects a portion of the two-dimensional FOV to a different angle, where they are optically reformatted to form one or multiple slits that are fed to a SG. The manufacturing of such a device for solar observations is difficult because the slices have to be thin ($<100 \mu m$) in order to achieve competitive spatial resolutions. In addition, a complex optical setup with several mirrors is required to re-image the many slices demanded to map a reasonable FOV. The limited FOV is usually fought using a complementary spatial scanning. A novel image slicing concept is being designed as part of the Multi-Slit Image slicer based on collimator-Camara (MuSICa\cite{calcines2012}) for the future, 4-m European Solar Telescope (EST\cite{collados2010,collados2013}), which differs from the already proven night-time solutions. MuSICa requires fewer reflections (e.g., three compared to the five of the Multi Unit Spectroscopic Explorer, MUSE\cite{laurent2006}), reducing instrumental polarization. In addition,  MuSICa has a symmetrical layout that simplifies manufacturing and alignment\cite{calcines2013}. A downscaled prototype of MuSICa is being tested using the GRIS spectropolarimeter (see Table \ref{tab:all_pol}) at the 1.5-m telescope GREGOR\cite{calcines2013b,calcines2014b}. For the final EST design, the MuSICa team foresees 8 identical slicers to map an 80 arcsec$^2$ FOV to 8 slits of 0.05$\times$200 arcsec$^2$ each. The slits will feed a single SG and the orthogonal polarization components of the resulting spectra, will be imaged in a single detector using a dual-beam configuration (see Sect. \ref{sect:mod}). To accomplish this, a first macro-slicer with eight 50-$\mu m$ slices (among the thinnest ever made\cite{calcines2013}) and a second micro-slicer with 16 slices are required. Two different image slicers designs have also been subject to feasibility studies for future solar space missions, see Refs. \citenum{calcines2014} and \citenum{suematsu2017}.

	\item \textit{Micro-lens array}: Another approach to snapshot spectroscopy is obtained by using a micro-lens array near the image plane, to create a sparse matrix of pupil images, one per lenslet. The void space in between the pupil images is filled with the individual spectra after a low-angle grating dispersion, without overlapping, provided that adequate filters are used. Such an approach was tried for solar observations in 1999 at the National Astronomical Observatory of Japan, see Ref. \citenum{suematsu1999}, using a 50$\times$50 array of 600 $\mu$m lenslets. The same team latter combined the micro-lens SG with crystal-based polarization modulators but reported a limited spectral performance and stray light issues\cite{suematsu2011}. These drawbacks were considerably improved by the design of MiHi (see Table \ref{tab:all_pol}). This instrument, currently under development, employs a different optical configuration which includes, among others, a second lenslet array and a stray light mask that are manufactured in the same substrate to avoid difficult co-alignments. MiHi has been used to do both single and dual beam (see Sect \ref{sect:spatiotemp_mod}) polarimetry with excellent results, e.g., reaching 0.3$\%$ polarimetric sensitivity after one second of integration with a spectral resolution of ~330000 and a fine spatial sampling of 0.065 arcsec/pixel. The micro-lens SG is a refractive instrument, thus is not achromatic and can produce fringes which demand a very stable setup to be properly calibrated\cite{jurcack2018}. 	  
\end{itemize}

%\begin{figure}	
%	\begin{center}
%		\begin{tabular}c
%			\includegraphics[height=10 cm]{mihi.png}
%		\end{tabular}
%	\end{center}
%	\caption[MiHi measurement example.]{MiHi measurement example TBC.}
%	\label{fig:mihi}    
%\end{figure} 

\section{Polarization modulators}
\label{sect:mod}

The basic working principle of imaging optical detectors is the photoelectric effect, this makes them primarily 
responsive to the intensity and wavelength of the incoming radiation. As a consequence, a necessary step to measure 
optical polarization is the encoding of the incoming Stokes vector into intensity images that can be registered using 
such detectors. This is the task of the polarization modulator and analyzer, namely, to perform the polarimetric 
mapping shown in Table \ref{tab:sp_map}. There are different techniques and components that can modulate spatially, 
spectrally and/or temporally the output intensity based on the values of the input Stokes parameters\cite{keller2015}. 
The latter can be retrieved post-facto by linearly combining the acquired intensity images using the modulator's 
response obtained in a separate polarization calibration procedure. There are various measurements and calibrations 
schemes, see e.g., Ref \citenum{deltoroiniesta2003}, that require specific data reduction procedures. The later can be 
complex due to the occasional non-linear entanglement of instrumental and ambient polarization effects, and is an 
ongoing research area, see e.g., Refs. \citenum{delacruzrodriguez2015,harrington2017, harrington2017b} and 
\citenum{harrington2018}. Polarimetry is thus reduced to precision differential photometry. At least four intensity 
measurements are required to estimate the complete Stokes vector. This is a fundamental aspect of polarimetric 
measurements because systematic, differential photometric errors produce spurious polarimetric signals. Measuring the 
full Stokes vector is desired not only from a sensing point of view but also for instrumental reasons. In the presence 
of instrumental polarization (produced by e.g., the telescope or adaptive optics system) the calibration of the data 
cannot be done without measuring the four parameters or having extra information, e.g., a given parameter is known to 
be zero a priori. In the pursuit of high-precision measurements, much effort has been put in reducing such effects for 
the different modulation techniques used in imaging solar polarimetry, this particular requirement has driven the main 
instrumental developments.

There are many ambient and instrumental phenomena that are time-dependent and can produce spurious signals (see Sect. \ref{sect:temp_mod}), additionally, some solar features can evolve rapidly (see Sect. \ref{sect:data_req}). Therefore, it is desirable to perform the polarimetric analysis as fast as possible. This can be attempted either with very fast ($>$kHz) temporal modulation, or with spatial or spectral modulation. Limitations in cadence and noise levels of the relatively large ($> 1$ Mpixel) imaging detectors required, have restricted the application of very fast temporal modulation. Full Stokes spatial or spectral technologies have not been successfully applied to do high-sensitivity solar polarimetry mainly due to current limitations in design and in the calibration of the separate measurement channels. By far the most employed technique used in ground and space based solar polarimetry (22 of 27 instruments in Table \ref{tab:all_pol}), is a combination of full-Stokes temporal modulation ($<100$ Hz) and partial spatial modulation (two channels) known as the dual beam setup, see Sect. \ref{sect:spatiotemp_mod}. This is because of its ability to partially reduce seeing and jitter induced artifacts, and the increased photon efficiency. Recent developments in polarization modulation techniques for solar observations mainly focus on increasing temporal modulation frequency, developing optical components for full-Stokes spatial modulation, increasing the spectral coverage while maximizing polarimetric efficiencies and testing newer devices for snapshot polarimetry such as polarization cameras or holographic elements. Extra details and examples are given in the following subsections.

\subsection{Temporal modulation}
\label{sect:temp_mod}

In a temporal modulation scheme, the modulator periodically changes its optical properties with time, to modify the polarization of the input beam and encode the Stokes parameters in fluctuations of the linear analyzer output intensity. All the instruments in Table \ref{tab:all_pol} make use of temporal modulation. Depending on the employed device, the modulator can change properties between the desired states in a continuous and smooth way, or in a discrete fashion. In any case the intensity value at each modulation state is registered using a scientific camera. Note that, even though the camera exposures have to be in phase to the modulation states, the actual detector readout could be done out of synchronization\cite{stenflo1994}. When doing one readout per modulation state, i.e., synchronous approach, the maximum modulation frequency (number of full-Stokes measurements per second) is limited by the maximum camera frame rate. In the asynchronous approach, frame exposure and readout are decoupled due to a special sensor design (see Sect. \ref{sect:detect}), allowing for frame rate and polarization modulation frequency to be independent.

The values of the Stokes parameters are retrieved by linearly combining the registered intensity images. As mentioned above, such an approach is sensitive to photometric effects varying on timescales faster or close to the modulation frequency. Since the frequencies used in modern polarimeters are higher than 1 Hz, see below, \textit{instrumental effects} related to thermal drifts and telescope configuration changes are commonly not a concern in this context. On the other hand, frame-to-frame camera artifacts and image jitter are typically an issue. The latter is generally related to the operation of Sun-tracking and adaptive optics systems or to uncontrollable vibrations in the telescope structure/building\cite{salmeida1992, salmeida1994, martinez-pillet2011}. The most relevant \textit{ambient effect} in ground-based systems is the spurious polarimetric signal introduced by atmospheric seeing, a.k.a. seeing-induced crosstalk, see e.g., Refs. \citenum{lites1987} and \citenum{salmeida1994}. The typical time scales of seeing evolution for daytime observations in competitive observing sites are of the order of 10 ms \cite{judge2004}, making this a suitable exposure time if post-facto image restoration wants to be used to reduce aberrations. It has also been shown using numerical simulations\cite{krishnappa2012} and measurements\cite{iglesias2016} that seeing-induced crosstalk considerably reduces, down to few times 0.01$\%$, for modulation frequencies above 100 Hz, and drops below the detection limit in the kHz regime\cite{casini2012}. 

\begin{figure}	
	\begin{center}
		\begin{tabular}c
			\includegraphics[height=8.5 cm]{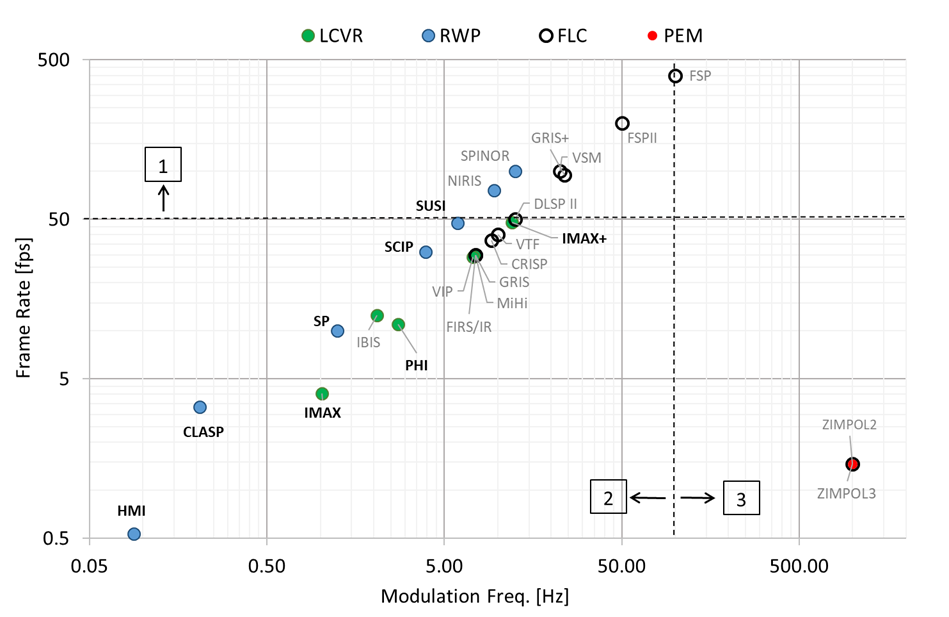}
		\end{tabular}
	\end{center}
	\caption[Detector frame rate vs. modulation frequency for the polarimeters listed in Table \ref{tab:all_pol}]{Detector frame rate (vertical axis) vs. modulation frequency (horizontal axis) for the polarimeters listed in Table \ref{tab:all_pol}. Each labeled dot represents a polarimeter, we use black labels for space, rocket and balloon borne instruments. The colors of the dots are used to highlight the different technologies employed for temporal modulation (see legend). The dashed lines separate different regimes relevant to ground-based observations. For frame rates above $\sim$50 fps (regime labeled 1), the image exposure time becomes well-suited for the implementation of post-facto image restoration, to reduce the aberrations produced by atmospheric seeing. Seeing not only reduces spatial resolution but also introduces polarimetric artifacts. For modulation frequencies below $\sim$100 Hz (regime labeled 2) these artifacts can be in the 1$\%$ level and thus most instruments employ a dual-beam configuration. For modulation frequencies well above $\sim$100 Hz (regime labeled 3), the polarimetric effects of the seeing are negligible. This regime has been explored by the ZIMPOL instrument which employs an asynchronous readout scheme, i.e., modulation frequency and detector frame rate are independent. All the other instruments have a synchronous readout and thus the modulation frequency is limited to typically a 4th (for crystal modulators) or an 8th (RWP's) of the detector frame rate. Note that both axes have a logarithmic scale. See the text for extra details.}
	\label{fig:mod_freq}    
\end{figure}

As a consequence of the above, instrumental developments using a temporal modulation scheme aim for a high modulation frequency to reduce jitter and seeing-induced artifacts. Most of the designs employ standard Charge Coupled Devices (CCDs) or Complementary Metal-Oxide Semiconductor (CMOS) architectures in a synchronous readout approach. Therefore, the availability of fast, low-noise cameras has been the bottle neck limiting the modulation frequency of high-precision polarimeters to below $\sim$100 Hz\footnote{We note that fast commercial cameras have been employed in high-flux applications, where detector readout noise is not a problem. Such as is the case of the high-cadence ($\sim$900 fps) polarimeter at the Japanese Solar Flare Telescope\cite{hanaoka2003,hanaoka2004}.}, see Fig. \ref{fig:mod_freq}. The dual beam setup employed by the majority of the current polarimeters circumvents the limitations in modulation frequency to some extent, see Sect. \ref{sect:spatiotemp_mod}. Two exceptions to the latter are the FSP, that employs a custom-made, low-noise 400-fps CCD detector in synchronous readout to reach 100 Hz modulation frequency; and ZIMPOL which is the only precision polarimeter that can achieve a very high frequency (in the 10 kHz range) by employing a specially masked CCD detector that allows asynchronous readout, see Sect.\ref{sect:custom}.

Retarders modify the polarization of an incoming beam by introducing a phase difference between the two orthogonal components of the electric field vector. Different kind of retarder-based solutions can be used to temporally modulate a beam. They differ mainly in their switching time (for electro-optical types), optical quality, available apertures, stability and operational spectral range. Common design issues among these are the following, see also Refs. \citenum{harrington2017} and \citenum{harrington2018}, (a) the presence of polarized fringes\cite{clarke2005}; (b) the variations of the retardance with beam angle of incidence and temperature\cite{hale1988}; and (c) the inability to keep the modulator response (quantified by its modulation matrix) constant across the desired spectral range, even when using a combination of achromatic or superachromatic retarders\footnote{The most common type are Pancharatnam configurations\cite{pancharatnam1955} made by combining different materials or several retarders of the same material at different orientations. More recently, they have been produced by stretching isotropic polymer foils\cite{samoilov2009}.}. In modern designs the latter issue is faced by giving up the modulation matrix achromatism, to obtain a simpler optical design that minimizes fringes and maximizes throughput among others, and performing a polarization calibration at each wavelength of interest. Once the desired accuracy of such a chromatic calibration procedure is achieved, the main design driver of the modulator becomes the optimization of the polarimetric efficiencies for the desired spectral range \cite{gisler2005}. The devices that have been used or are being tested for solar spectropolarimetry are described below.
 
\begin{itemize}
	\item \textit{Rotating Waveplate} (RWP): This solution employs an electric motor to rotate a waveplate and produce a smooth variation of its fast optical axis orientation. The main advantage is that waveplates with highly-customizable retardances can be manufactured to work in most of the optical spectrum, with high surface quality and homogeneous properties across their clear aperture, and that they are stable in time\cite{keller2015}. RWP's typically present a broader working spectral range, up to few hundred nm, and can be used in spectral windows where liquid crystal (LC) solutions are not suitable, particularly in the UV regime below 400 nm. On the other hand, RWP's have characteristic disadvantages\cite{wachter2012}: (a) the moving parts imply a more complex instrument design due to the increased mass, power consumption and vibrations; (b) the mechanical rotation plus unavoidable misalignments and residual shape errors in the waveplate, introduce a variable beam synchronous to the modulation\cite{stenflo1994}; (c) non-uniformities in the rotation; and (d) the modulation frequency is mechanically limited by the retarder rotating speed to below $\sim$25 revolutions s$^{-1}$\cite{hanaoka2012}. Both (b) and (c) render the RWP prone to crosstalk errors and require elaborate and costly engineering.  Image stabilization techniques and high-quality RWP units have been produced for reducing wobbling and jitter, this is particularly the case for space and rocket based instruments, where this technology has achieved the highest readiness levels, e.g., see SP or HMI in Table \ref{tab:all_pol}. For example, the modulator employed in the recent sounding-rocket CLASP (see Table \ref{tab:all_pol}), employs an actively-controlled RWP unit that can damp non-uniformities and wobbling to reduce polarimetric errors down to the 0.01$\%$ level\cite{ishikawa2016}. For ground-based systems, where design constraints are more relaxed, budget is generally lower and higher modulation frequencies are required to reduce seeing effects, RWP's are typically replaced by the more convenient electro-optical modulators, whenever the targeted spectral range allows it.
   
    \item \textit{Liquid Crystal Variable Retarder} (LCVR): A thin layer of birefringent LC contained between two glass plates and a pair of electrodes, can be used to change the polarization of a beam propagating perpendicular to the plates. By applying a voltage to the electrodes, either the aspect ratio or the orientation of the individual crystals can be controlled, producing a modification of the device retardance or its optical axis orientation. The first kind are called LCVR's with the most commonly type employed being nematic LC's. Crystal-based modulators avoid the issues introduced by moving parts and can reach higher modulation frequencies compared to RWPs. On the other hand, large devices are difficult to produce and their optical properties can considerably change with temperature and across their clear aperture. LCVR's have traditionally been the slowest of the LC's used in solar polarimetry with switching times of the order of 10 ms.  Moreover, the switching of these crystals is asymmetric, i.e., it takes longer time to switch from low to high retardance, which is done by removing the supply voltage. Higher voltages can be used to overdrive nematic LC's and reduce their switching time\cite{thalhammer2013}, this at the expenses of an increase in thermal sensitivity \cite{delacruzrodriguez2015}. Recent improvements in LCVR's manufacturing are bringing their switching time down to the ~1 ms level\cite{snik2014}. LCVR's can be generally manufactured with larger apertures than other LC's ($\sim$50 mm) and can be made more achromatic if two crystal layers are combined\cite{capobianco2012, deltoroiniesta2012}. LCVR's were successfully employed in the ballon-borne IMAX instrument, and have been recently space qualified\cite{heredero2007,uribe2011} and included in the PHI spectropolarimeter on board of the Solar Orbiter mission, see Table \ref{tab:all_pol}. Other recent LCVR developments in solar polarimetry include studying devices with different working principles, such as dual-frequency LC's (see below).
   
    \item \textit{Ferro-electric Liquid Crystal} (FLC):  FLC's can also be confined in thin layers to produce a device with electrically controllable polarization properties. The orientation of the FLC's fast optical axis rotates between two bistable positions when the applied voltage switches polarity\cite{gandorfer1999}. The typical switching angle  is 45$^\circ$ and the retardance, controlled by the plate thickness, half the central working wavelength. Two FLC's can be used to produce the four modulation states required to measure the complete Stokes vector. FLC's switch significantly faster ($\sim$100 $\mu s$ level) than other LC's, two or more devices can be combined to work in the 400 to 1700 nm wavelength range approximately and are a very commonly used solution in ground-based solar polarimeters, e.g., 13 of the 19 ground-based instruments in Table \ref{tab:all_pol} use FLC's. The main drawbacks of FLC's are the strong thermal sensitivity of their retardance and switching angle, demanding an accurate ($\sim$0.1 $^\circ C$) temperature control in any high-precision ($\sim$0.01$\%$) application; the strong dispersion of their retardance\cite{gisler2003}; and that they are limited to small apertures (typically of order 10-40 mm) mainly due to their vulnerability to mechanical stress. If high polarimetric efficiencies are required in a broad spectral range, the strong dispersion is handled by combining FLC's with static retarders to minimize the spectral dependence of the efficiencies. This procedure was introduced in Ref. \citenum{gisler2005} and applied in e.g., the FSP, see Table \ref{tab:all_pol}.
   
   	\item \textit{Piezo Elastic Modulator} (PEM): PEM's are based on glass that becomes birefringent under mechanical 
   	stress\cite{stenflo1985}. A standing acoustic wave is created using piezo-electric transducers at the resonance 
   	frequency of the glass plate, producing an equally rapid modulation of its retardance. They can be used at 
   	wavelengths that range from the vacuum UV to the IR. However, given that the crystal size defines the resonance 
   	frequency, only very high modulation frequencies can be used (20 to 50 kHz) for practical PEM apertures of the 
   	order of several 10 mm, ruling out any application with synchronous camera readout\cite{stenflo1994}. Additionally, 
   	no full-Stokes modulator has been developed due to the practical difficulties of the required phase locking of two 
   	PEM's\cite{gandorfer1999}. Another type of resonant temporal modulator, similar to PEM's, was foreseen using 
   	Pockels and Kerr cells. However, partially due to the small apertures available such technology did not see 
   	widespread application in solar polarimetry\cite{gandorfer1999, gandorfer2004}.
  
  	\item \textit{Dual-frequency Liquid Crystal} (DFLC): DFLC's are a kind of LCVR that are under study to be used in solar polarimetry\cite{nagaraju2018}. The retardance of a DFLC changes between a very low (ideally 0) and a manufacturing-tunable value (e.g., half wavelength), when the frequency of the applied voltage surpasses a critical number\cite{mrukiewicz2015}. Initial studies suggest that two DFLC's can be combined with two static retarders to produce a full-Stokes modulator with achromatic polarimetric efficiencies in the 600 to 900 nm range. The main advantage of DFLC's is that they can switch as fast as FLC's\cite{golovin2003} but can be produced with larger apertures. The latter is relevant for application at future large-aperture solar telescopes.
\end{itemize}

\subsection{Spatial modulation}
\label{sect:spatial_mod}

The Stokes vector can also be modulated in space using devices that split the incoming light into spatially separated polarimetric channels. Such an approach can be used to measure multiple Stokes parameters simultaneously, by linearly combining the intensity signals registered in the different channels. With 2 or 3 channels, only specific components of the seeing and jitter induced artifacts are reduced, see Sect \ref{sect:spatiotemp_mod}. On the other hand, if the full Stokes vector is instantaneously sampled (minimum four channels), these artifacts vanish, see Sect. \ref{sect:temp_mod}. Similar to the spectroscopic case, spatial modulation has also the benefits of reducing artifacts and smearing due to the evolution of solar signals, and of increasing SNR by making use of all the available photons reaching the modulator at a given time. Spatial modulators, however, are susceptible to differential effects among the polarimetric channels. Although imaging, full-Stokes, spatial modulators have been developed using a variety of solutions, such as prism-based, 4-way splitters for the Visible and IR\cite{compain1998,devogele2017}; a Fourier-transform modulator based on calcite wedged crystals\cite{oka2003}; or a 6-way splitter based on a Wollaston prism array\cite{mu2015}; they have not been definitively demonstrated in a competitive solar polarimeter. This is due to their apertures, differential wavefront aberrations, throughput and polarimetric efficiency figures, which are not compliant with the required spatial resolution, FOV and sensitivity levels compared to the well-proven dual-beam solution, see Sect. \ref{sect:spatiotemp_mod}. Note that a full-Stokes, spatial modulator with the required optical performance is highly desirable because in combination with existing integral field spectrometers, as done for night'time astronomy\cite{rodenhuis2014}, it can be used to do high-resolution, snapshot solar spectropolarimetry, see Table \ref{tab:sp_map}. Recent relevant devices developed for spatial modulation are high-quality micropolarizers and microretarders grids, attached in front of imaging detectors to produce polarization cameras (see Sect. \ref{sect:detect}), and polarization gratings, see Sect. \ref{sect:sp_mod}.

%A candidate design that has been recently studied for solar applications\cite{feller_PC2018} involves utilizing a 6-way polarizing beam splitter made of XXX, that only employs internal reflections to avoid ...  XXXXALEX and MICHIEL DESIGN, SHALL WE INCLUDE A SKETCH ???XXX. Other recent relevant developments for spatial modulation are high-quality micropolarizers and microretarders grids, attached in front of imaging detectors to produce polarization cameras (see Sect. \ref{sect:detect}), and polarization gratings, see Sect. \ref{sect:sp_mod}.

\subsection{Spatio-temporal modulation}
\label{sect:spatiotemp_mod}

The most widespread type of polarization modulation used in solar instruments is spatio-temporal, in the form of a dual beam setup. A device that can split the beam in its two orthogonal polarization components is used as linear analyzer and located after the temporal modulator (on of the types discussed in Sect. \ref{sect:temp_mod}), see e.g., Refs. \citenum{lites1987} and \citenum{collados1999}. Two detectors, or two sections of the same detector, are used to image both orthogonal beams. The combination of two simultaneous measurements, either by doing joint or independent demodulations\cite{lacasse2011}, eliminates the most important component of jitter and seeing-induced artifacts\cite{collados1999,casini2012}, namely the crosstalks from Stokes I to Q, U and V. The main factors limiting the accuracy of the dual beam technique are optical differences among the two beam paths, difficulties in the beams co-registration and the inability to calibrate the detectors differential gain tables and non-linearities below the $\sim 0.1\%$\cite{collados1999,snik2014} level, see Sect. \ref{sect:detect}. Many improvements have been made in birefringent materials, in the beamsplitting crystal cement layers and anti-reflective coatings, to produce two-way polarizing beam splitters that have high extinction ratio, transmission and surface quality, in sufficiently large spectral windows from the UV to the IR\cite{dobrowolski2001}. Most of the beam splitters used in solar polarimetry are different realizations of beam displacers\cite{semel1987}, beam splitter cubes and Wollaston prisms. They are optimized for each specific instrumental setup including the required spatial splitting (related to the detector size and optical layout), wavelength range and extinction ratio. Table \ref{tab:beamsp} lists the most commonly used devices along with other interesting custom designs or assemblies.

\begin{table}[!htp]
	\caption[Polarizing beamsplitters used in solar spectropolarimetry.]{Polarizing beamsplitters used in solar spectropolarimetry. For each solution the table gives the device name, a short comment or description, and example instruments. The corresponding references in Table \ref{tab:all_pol} can be consulted for further details on each technology. The customs designs were specially developed and demonstrated for the specified instrument.}
	\centering
	\small
	\begin{tabular}{m{3cm} m{9.5cm} m{2.5cm}}
		\noalign{\smallskip}     		
		\cline{1-3}  
		\noalign{\smallskip}  
		Device & Comments / Description & Instrument   \\
		\noalign{\smallskip}  
		\cline{1-3}	
 		\noalign{\smallskip}  	
        \multicolumn{3}{l}{\textit{General purpose designs}}\\	
		\noalign{\smallskip}  
		\cline{1-3}
		\noalign{\smallskip} 	       		
		Wollaston Prism & The output beams are refracted at nearly opposite angles. Broadly employed in night-time polarimetry but not in solar applications. Used near a pupil plane which increases differences between the optical paths\cite{gandorfer2002b, beck2010}. & FIRS/IR, VIP  \\	
		\noalign{\smallskip}  
		\cline{1-3}	
		\noalign{\smallskip} 	       				
		Modified Savart Plate & The two refracted output beams are parallel. Typically located close to the detector. Widely used in solar polarimeters. & IBIS, SP, DLSPII, VSM  \\	
		\noalign{\smallskip}  
		\cline{1-3}	
		\noalign{\smallskip} 	       				
		Wire-grid polarizer &  Based on a wire-grid linear polarizer, implemented with thin-film technology\cite{bischoff2018}. It has lower performance compared to crystal-based devices, for narrow band and/or photon-starved applications. & SPINOR  \\	
		\noalign{\smallskip}  
		\cline{1-3}		
		\noalign{\smallskip} 	       								
		Beam splitter cube &  The two beams output at a large angle, e.g., 90$^\circ$, and imaged in two separate detectors. Broadly used in solar, including space-born, polarimeters. & IMAX, HMI, CRISP, SUSI, SCIP  \\	
		\noalign{\smallskip}  
		\cline{1-3}		
		\noalign{\smallskip}  	
        \multicolumn{3}{l}{\textit{Custom designs / assemblies}}\\	
		\noalign{\smallskip}  
		\cline{1-3}		
		\noalign{\smallskip} 	       				
		5-Cubes & Five thin-film, beam splitter cubes are cemented together to produce the splitting\cite{collados2007}. The output beams are parallel and travel identical optical paths within the device & GRIS  \\	
		\noalign{\smallskip}  
		\cline{1-3}		
		\noalign{\smallskip} 	       				
		Double Wollaston & Two Calcite Wollaston prisms are combined, the first one acts as beam splitter while the second one produces two parallel beams at the output. & NIRIS\\	
		\noalign{\smallskip}  
		\cline{1-3}		
		\noalign{\smallskip} 	       				
		Normal refractions& The beam displacement is accomplished by using
		four cemented crystal pieces made of fused silica\cite{feller_PC2018,vannoort_PC2018}. Wavelength dependent refraction is completely eliminated in this design due to normal incidence and exit angles. Both beam paths are identical& FSPII\\	
		\noalign{\smallskip}  
		\cline{1-3}		
		\noalign{\smallskip} 	       				
		Broadband & A first beam splitter cube produces two orthogonal beams which are then subject to two mirror reflections each, to make them parallel. After traveling the same path, both beams are reimaged to a single detector using relay lenses. & ViSP\\	
		\cline{1-3}		
		\noalign{\smallskip} 	       				
		Ultra Violet & The beamsplitting is accomplished by two reflections on two $MgF_2$ birefringent plates placed at the Brewster angle. & CLASP\\	
		\noalign{\smallskip}  
		\cline{1-3}	 		
	\end{tabular}
	\label{tab:beamsp}
\end{table}

An alternative technique used to do spatio-temporal modulation is the beam exchange\cite{donati1990}, which is similar to the dual beam except that it requires an extra measurement. The latter is acquired after exchanging the beams of the two channels using e.g., a rotating half-wave plate as done in the first solar application, see Ref. \citenum{bianda1998}. The four acquisitions are latter combined to reduce both crosstalk from Stokes I to Q, U and V, and the artifacts produced by differential sensor gain tables. The beam exchange is commonly used in night astronomy polarimeters (e.g., Ref. \citenum{rodenhuis2012}) and was further explored in solar observations with promising results \cite{bommier2002}. However, it was never broadly adopted. The extra measurement required makes the SNR vs. spatial resolution trade off even worse, see Fig \ref{fig:tihrs}.

\subsection{Spectropolarimetric modulators}
\label{sect:sp_mod}

The development of devices with the ability to perform both the spectral and polarimetric analyses simultaneously, see Table \ref{tab:sp_map}, has experienced a strong growth in the last 10 years resulting in various original instruments for astronomy and remote sensing\cite{hoeijmakers2016,snik2009,vanharten2011,oka1999,oka2003,sparks2012,oka2006}. These novel concepts have not seen similar proliferation in solar polarimetry, mainly because the realization of a system that can image an extended solar target with competitive spectral and spatial resolution, FOV and polarimetric sensitivity has not been demonstrated. Two popular techniques are (a) \textit{Channeled polarimetry}\cite{alenin2014, sparks2012} which can be used to encode the input Stokes parameters either into intensity variations of the output spectrum\cite{snik2009,oka1999} or into a set of spatial fringes\cite{oka2003}. The polarization information is commonly retrieved via reconstruction algorithms that involve fitting the imaged observables. And (b) \textit{Polarization holography} which utilizes gratings with anisotropic profiles that can be tuned to disperse the beam at different orders depending on its polarization state\cite{kakauridze2006,attia1983}. \textit{Polarization gratings}\cite{packham2010} sensitive to linear polarization only present high efficiency ($\sim 99\%$\cite{oh2008}) in a broad spectral range, and has been used as analyzers in astronomy\cite{davis2001}. The combination of several polarization gratings has been recently used to form a full-Stokes modulator known as \textit{polarization holographic element\cite{kakauridze2008}}, and applied in a solar polarimeter, see Refs. \citenum{kvernadze2017} and \citenum{kvernadze2017b}. The performance of the reconstruction algorithms, along with the extra demands imposed to the spectral resolution and range; and the inability to image extended sources without extra spatial scanning\cite{hoeijmakers2016}; have diminished the implementation of channeled polarimetry and polarization holography to solar observations so far.

\section{Imaging detectors for high-precision polarimetry}
\label{sect:detect}

The detector used to register the modulated intensity signal is a crucial component of any imaging solar polarimeter\cite{stenflo1994}. The great majority of instruments use standard CCD or CMOS sensors, although the latter are preferred in recent designs. In spite of this, the notable success of the specially-modified detector used in ZIMPOL, and recent advances in polarization cameras and active pixel sensors motivate also the development of custom solutions. Due to the differential nature of the polarization measurements and low photo-electron counts obtained in high-resolution solar applications, there are specific aspects of the detectors that gain relevance compared to spectroscopic or broadband imaging applications, see e.g., Ref. \citenum{iglesias2016}. Some of these are described in the following subsections, within the context of each detector technology.

\subsection{CCD detectors}
\label{sect:ccd}
	
 CCDs, see e.g., Ref. \citenum{janesick2001}, were the first kind of imaging detectors adopted for solar polarimeters\cite{stenflo2017}. Given the architecture and working principle of CCDs, most notably the absence of readout transistors within the pixel structure, they have historically offered a more linear and spatially homogeneous response than the later-introduced CMOS\cite{janesick2006,janesick2007,janesick2013}. However, sharing a single charge readout amplifier among many pixels strongly limited the maximum achievable frame rate, particularly for large detectors. This was a major drawback when trying to employ CCDs for synchronous detection in a temporal modulation scheme (see Sect. \ref{sect:temp_mod}), and what motivated the development in the 1990's of the ZIMPOL concept for ground-based solar observations, see below. CCDs are of wide-spread use among modern solar polarimeters of all platforms (13 of 27 instruments in Table \ref{tab:all_pol} use CCDs), with the main architecture of choice being frame-transfer to maximize duty cycle (critical in photon-starved conditions such as solar spectropolarimetry). Modern scientific CCDs present low readout noise figures (order of 1 to 50 $e^-$ RMS), important to minimize integration time in photon-starved conditions; high quantum efficiency, mostly in silicon-sensitive spectral ranges (from $\sim$300 to $\sim$1100 nm plus the X-ray); and large full well capacity (order of 300 $ke^-$). Relevant calibration issues when applying them to polarimetry are common-mode noise\cite{janesick2001} and frame-to-frame variable offsets, which can be reduced by subtracting the signal of specially shielded pixels\cite{iglesias2016}; response non-linearities\cite{keller1996}, that can be calibrated down to below $\sim 1\%$ using look-up tables\cite{collados2007}; and, for shutter-less designs, frame-transfer artifacts that can be numerically corrected\cite{iglesias2015}. The need for large sensors areas, to properly sample the focal plane of bigger telescopes; for higher frame rates; and the great performance improvements and cost reductions, have motivated the usage of CMOS solutions in many recent instruments, i.e., all polarimeters dated 2017 or beyond in Fig. \ref{fig:all_pol_aperture} employ CMOS detectors.

\begin{figure}	
	\begin{center}
		\begin{tabular}c
			\includegraphics[height=8.5 cm]{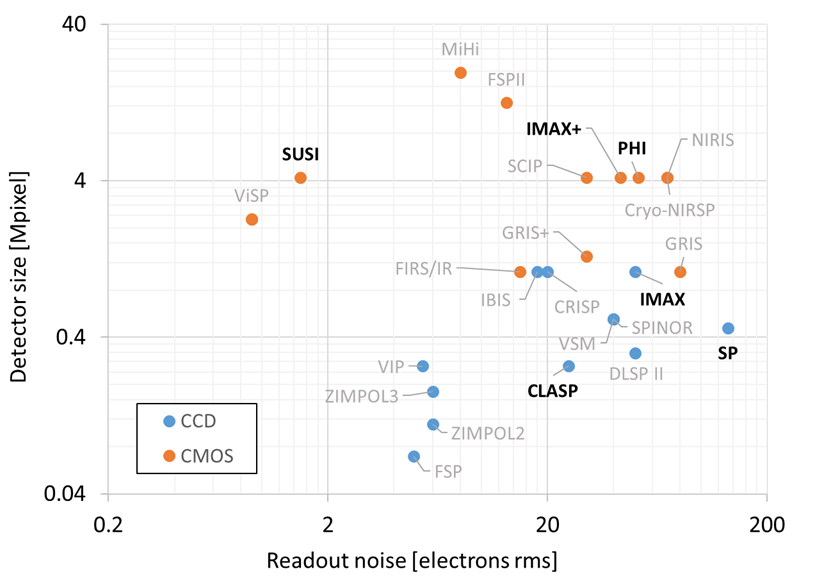}
		\end{tabular}
	\end{center}
	\caption[Detector size vs. readout noise for most of the polarimeters listed in Table \ref{tab:all_pol}.]{Detector size (vertical axis), readout noise (horizontal axis) and technology (colors, see legend) for most of the polarimeters listed in Table \ref{tab:all_pol}. We use black labels for rocket, space and balloon borne instruments. Note that both axes have a logarithmic scale.}
	\label{fig:detectors}    
\end{figure} 
	
\subsection{CMOS detectors}
\label{sect:cmos}

CMOS detectors\cite{janesick2002} are characterized for having the charge to voltage amplifier (and sometimes signal-processing circuitry) within each individual pixel. This imposes size constraints which result in performance and calibration limitations\cite{janesick2006} that traditionally diminished the application of CMOS to high-sensitivity polarimetry. Partially driven by the mass consumer electronics market and industrial applications, the performance and diversity of CMOS detectors has radically improved in the last decade\cite{bigas2006,waltham2010}. The most relevant aspects that have motivated the usage of CMOS sensors in recent ground and space based solar polarimeters are the following. (a) Correlated multiple sampling and other techniques have strongly reduced readout noise levels to $\sim2 e^-$ RMS or even lower\cite{guidash2016}, see Fig. \ref{fig:detectors}. If large well-capacity and frame rate are used then common figures are $\sim40 e^-$ RMS. (b) Many detectors provide, in addition to the traditional rolling-shutter, a global-shutter readout scheme, which allows exposing all the pixels simultaneously and while the charges of the previous exposure are being read. This increases duty cycle to practically $\sim$100$\%$ and is particularly beneficial in temporal polarization modulation\cite{piqueras2013}. (c) Further miniaturization of gate structures, the usage of micro lens arrays and developments in back-side illuminated devices have substantially increased quantum efficiency and filling factors in CMOS detectors to make them comparable to CCDs. Moreover, hybrid CMOS detectors can function in IR and UV bands where CCDs are not sensitive. Three examples from Table \ref{tab:all_pol} are the nitrogen-cooled HgCdTe cameras (sensitive up to 10 $\mu m$) used in GRIS; the InGaAs cameras used in GRIS+\footnote{InGaAs cameras are also used in the Japanese Solar Flare Telescope \cite{sakurai2018}.}, which present efficiencies above 70$\%$ for 1000-1700 nm and low dark current using only thermo-electric coolers; and the back-illuminated, thinned UV detectors selected for SUSI, which have $\sim$80$\%$ efficiency at 250 nm. (d) CMOS detectors are nowadays produced with sufficient homogeneity and fewer cosmetics defects for sizes up to 20 Mpixel. (e) Multiple parallel outputs (typically one or two per column) and higher readout frequency allow for high cadences with low noise, e.g.,  the SUSI UV detector can reach 48 fps with 1.4$e^-$ RMS read noise in a 4 Mpixels format, while GRIS+ InGaAs camera can run at 95 fps with a 1.3 Mpixel resolution. (f) Current CMOS manufacturing technology can produce pixels that are small (5 to 10 $\mu m$) helping to produce more compact and simpler instruments that minimize instrumental errors. (g) CMOS sensors have been proven resistant to radiation harness and have been space qualified to be used in solar polarimeters, e.g as in PHI\cite{piqueras2013}, see Table \ref{tab:all_pol}. (h) Response non-linearities are particularly detrimental in polarimetry\cite{keller1996}. Due to the coexistence of charge well and amplifiers within the same pixel structure, CMOS sensors response is more non-linear than that of CCDs. Due to this, special characterization techniques have been developed to e.g., measure conversion gain\cite{bohndiek2008b} and quantum efficiency\cite{fowler1998}. Non-linearities have been calibrated below the $1\%$ level in some cases\cite{wachter2012}, e.g., for HMI (see Table \ref{tab:all_pol}). However, the accuracy of such a calibration can be strongly sensor-dependent due to the specific hardware solutions implemented by the manufacturers for readout and power supply. For example the CMOS detectors used in MiHi, see Table \ref{tab:all_pol}, present a non-linear dependence of the pixel signal with the illumination level of the corresponding row, due to deficiencies in the shared ramp analog to digital converter\cite{sant_PC2018}. In addition to the non-linearity issue, CMOS present other disadvantages including small well depths, typically in the ~30 $ke^-$ range, and they have not been fully demonstrated suitable to do very high sensitivity ($10^{-4}$ or below) polarization measurements of the Sun.
	
\subsection{Custom detectors}
\label{sect:custom}

The most successful custom-sensor developed for solar polarimetry was done for ZIMPOL\cite{povel1990}. ZIMPOL uses a \textit{specially-masked CCD detector} with three out of four rows covered. In combination with a synchronous charge shifting, the mask allows the separate accumulation of the photo-charges corresponding to the 4 different modulation states. The accumulation typically includes a large number of modulation cycles between subsequent frame readouts\cite{gandorfer1997}. ZIMPOL allows for fast modulation up to some 10 kHz (in combination with PEMs as polarization modulators, see section 3.1), virtually eliminating seeing and jitter induced artifacts and reaching $10^{-5}$ polarimetric sensitivity. Further, the decoupling between modulation frequency and frame rate allows to accumulate a large number of photo-charges, corresponding to a significant fraction of the full well, and thus to mitigate the effect of readout noise. However, the ZIMPOL approach also presents drawbacks, in particular when high spatial resolution is required\cite{iglesias2016}. These are the non-square pixels resulting from the usage of a micro-lens array, required to maximize filling factor, which produce two different spatial sampling frequencies that complicates image restoration; up to date the ZIMPOL concept has been implemented in slow readout sensors ($\sim$2 fps) limiting the study of fast events and the application of image restoration techniques in combination with high duty cycle. The FSP, employs a frame-transfer, fully-depleted CCD detector\cite{ihle2008,ihle2012} that was custom-made with column parallel readout. The 400 fps, almost 100$\%$ duty cycle and low noise (4.9 $e^-$ RMS) of the detector allowed FSP to avoid seeing effects  and reach the $\sim$0.01$\%$ polarimetric sensitivity level using FLC-based temporal modulation only\cite{iglesias2016}. However, the sensor was small (264x264 pixel$^2$) and the development of a full-sized version (1024x1024 pixel$^2$) proved difficult and expensive when compared with CMOS competitors. The prototype of another very promising custom detector type for polarimetry has developed very recently: \textit{the Quadropix DePFET}\cite{baehr2017,baehr2018}. It involves using a Depleted p-channel Field-Effect Transistor (DEPFET) active pixel sensor, that consists of four sub-pixel structures for each individual pixel. The sub-pixels can be controlled such that only one accumulates all the photo-charges generated in the pixel during a time interval that corresponds to a given modulation state. The switching time between sub-pixels is extremely short (order 10 ns) allowing a combination with fast modulators such as PEM's or FLC's. This solution has the same benefits than the ZIMPOL approach but avoids sensor masking and the usage of micro-lens array. In contrast to the current version of ZIMPOL the DePFET sensor technology employs fast column-parallel readout similar to the CCD detector type used in FSP. With the present VERITAS readout ASICS\cite{porro2013} available for the DePFET sensors, frame rates of order 100 fps will be possible for 1 Mpixel sensors. The noise properties of the DePFET Quadropix are on the same excellent level as the CCD used in FSP, mainly because of the very similar readout electronics architecture.

\subsection{Polarization detectors}
\label{sect:pol_det}
Another custom detector design for polarimetry that has seen important developments in the last decade, due to the rapid progress in microlithography, are polarization detectors\cite{}. These can be manufactured using pixel-size micropolarizers or microretarders in front of the imaging detector, which is accomplished by writing sub-wavelength periodic structures that work as wire-grid polarizers or retarders on the semiconductor material. Polarization detectors sensitive only to linear polarization have proven to deliver 0.3$\%$ sensitivity in astronomical applications\cite{vorobiev2018}. Moreover, due to their dramatically smaller size, lower power consumption, mechanical robustness and snapshot-capabilities, among others, polarization cameras are ideally suited for space, balloon or rocket applications. They have been recently used to explore visible coronal emission lines during the 2017 total solar eclipse in the US, mainly as a demonstrator for a balloon-borne coronagraph\cite{tomczyk2017,burkepile2017} currently under development. Polarization detectors employ spatial modulation and thus different neighboring pixels have to be combined to retrieve the Stokes parameters, e.g., four to measure Stokes I, Q and U \cite{nelson2017}. This has two main effects, firstly, the FOV and/or spatial resolution are restricted. Secondly, the polarimetric sensitivity is limited by the ability to calibrate the differential effects among the combined pixels, e.g., optical aberrations, pixel point spread functions, etc. Full-Stokes polarization cameras are typically obtained using two detectors, one of which is combined with a retarder to be sensitive to circular polarization\cite{tu2017}; or using a single detector and a crystal-based, temporal modulator\cite{vedel2011}. To the best of our knowledge these have not been used in solar applications yet. 

\section{Summary}

In this review we have described the main technology used in state-of-the-art solar spectropolarimeters developed in the last two decades, see Sect. \ref{sect:opt_pol}. An emphasis was made on full-Stokes, optical instruments that aim to obtain the challenging high-resolution, high-sensitivity data demanded by many important open science questions (see Sect. \ref{sect:data_req}). We have also included some instrumentation and technological concepts under development to provide an outlook on promising future design directions, particularly in the light of the upcoming, large-aperture solar telescopes. We summarize below the most relevant points arising from each section of the manuscript.

\textit{Wavelength discriminators:}
\begin{itemize}
	\item State-of-the-art polarimeters employ multiple Fabry-Perot filtergraphs or echelle spectrographs, which both are mature technologies that can adequately sample most of the targeted solar spectral lines. Recent efforts focus mostly on increasing efficiency in the UV and IR spectral regimes and observing many spectral lines simultaneously (see Sect. \ref{sect:spect}) with increased spatial resolution.
	\item A technique for post-facto image restoration of spectrograph data has been recently demonstrated. It considerably improves the achievable spatial resolution of slit-spectrograph scans across a 2D solar image (see Sect. \ref{sect:sg}).
	\item Lithium niobate Fabry-Perot interferometers have been space qualified and included in the PHI instrument, bringing an alternative to the Michelson-based approach used in MDI and HMI (see Sect. \ref{sect:fg}). 
	\item Five integral field solutions are being developed and have been tested for high-resolution solar polarimetry. These devices can perform snapshot-spectroscopy, thus improving data simultaneity and the overall SNR when imaging extended sources, and are a priority for large aperture solar observatories. The resulting spatial and spectral FOV is limited requiring a complementary spatial scanning system for large (e.g., sunspots) targets (see Sect. \ref{sect:ifs}).
	\item The challenging IR spectral domain (useful to study the important chromosphere and transition region) will be further explored by the 4-m DKIST with great sensitivity up to 5 $\mu m$. The UV regime is still poorly explored in terms of spatially resolved spectropolarimetric measurements. The rocket-based CLASP has performed such measurements in the Ly alpha 121 nm line. The upcoming balloon-borne SUSI is designed to explore the 300-400 nm band (see Sect. \ref{sect:spect}).
\end{itemize}

\textit{Polarization modulators:}
\begin{itemize}
	\item  Temporal modulation using synchronous readout is still limited to seeing-vulnerable frequencies ($\sim$100 Hz), by the detector frame-rate and readout-noise figures (see Sect. \ref{sect:temp_mod}). As a consequence, the great majority of high-sensitivity, current and upcoming instruments use a dual-beam configuration to reduce the main component of seeing and jitter induced artifacts, namely measurement errors in form of crosstalk from Stokes I to Stokes Q, U, V; and to improve photon efficiency (see Sect. \ref{sect:spatiotemp_mod}). Some exceptions are ZIMPOL, FSP and few fast instruments found in high-flux applications (see Sect. \ref{sect:detect}). The dual beam technique is still mostly limited to $\sim 0.1\%$ sensitivities by differential effects among the channels, including optical aberrations and camera stability or calibration issues.
	\item The most common technology for ground-based, temporal modulation in the visible and IR spectral bands are crystal-based. FLC's are used, whenever their limited aperture permits, because they are faster than the alternative LCVR's. For the upcoming large-aperture telescopes this is an issue and thus DFLC's are being explored. They are as fast as FLC's and can be produced with larger apertures (see Sect. \ref{sect:temp_mod}).
	\item LCVR's have been recently space qualified and included in the PHI instrument. This brings an alternative to the mature and space-proven RWP's, although within a more limited spectral range (see Sect. \ref{sect:temp_mod}). 
	\item To date, Full-stokes spatial modulators with the required performance have not been developed yet. When combined with integral field solutions, they are a promising alternative to do high-resolution snapshot solar spectropolarimetry (see Sect. \ref{sect:spatial_mod}).
	\item There are many different types of polarizing beam splitters used in dual beam setups, depending on wavelength range, polarimetric sensitivity and optical setup (near pupil vs near a focal position). Polarization gratings are now an alternative although untested in a solar application (see Sect. \ref{sect:spatiotemp_mod}).
	\item Two novel techniques for imaging spectropolarimetric modulation have recently produced interesting results in astronomy, i.e., channeled polarimetry and polarization holography, although with limited performance compared to the requirements in high-resolution solar applications. Only the latter has been tested in a full-Stokes solar polarimeter (see Sect. \ref{sect:sp_mod}).
\end{itemize}

\textit{Imaging detectors:}
\begin{itemize}
	\item CMOS detectors have made great improvements in the last decades, mainly in terms of cost reduction, noise, frame rate, detector size and stability. Moreover, their increasing usage has pushed back CCDs in many upcoming instruments. However, careful calibration, which can be strongly manufacturer-dependent, is required to reduce non-linearities and crosstalk, among others (see Sect. \ref{sect:ccd} and \ref{sect:cmos}).
	\item Hybrid CMOS detectors dominate IR applications with InGaAs cameras now offering competitive cadence and noise figures with convenient thermo-electric cooling. In the UV, back illuminated CMOS detectors can reach high quantum efficiency levels while offering frame rates in the thens of Hz range and few-electrons noise (see Sect. \ref{sect:cmos}).
	\item The ZIMPOL solution is still the only high-sensitivity instrument that can consistently operate in the kHz and 10 kHz regime thanks to a custom-made sensor design. Other similar approaches are under study based on DEPFET sensors (see Sect. \ref{sect:custom}).
	\item Polarization cameras have made great improvement in recent years and have been proven in astronomy to measure linear polarization down to the 0.3$\%$ sensitivity level. They were also tested in a ground-based solar coronagraph with promising results. They are small and convenient for space, rocket and balloon applications. However a full-Stokes version requires extra components, they impose a constraint in resolution and FOV, and they are prone to crosstalk due to differential spatial response (see Sect. \ref{sect:pol_det}).
\end{itemize}
	
\acknowledgments 
We thank the three anonymous referees for their valuable help to improve the quality of this review. We are also grateful to the experts of the described instruments for their crucial information, in particular to M. Van Noort, H.P Doerr, F. Zeuner, K. Nagaraju and K. Sant. FAI acknowledges a fellowship of CONICET and appreciates support from UTN projects UTI4035TC and UTI4915TC.

%%%%% References %%%%%

\bibliographystyle{spiejour}   % makes bibtex use spiejour.bst
\bibliography{../../iglesias}  % bibliography data in report.bib

%%%%% Biographies of authors %%%%%

\vspace{2ex}\noindent\textbf{Francisco A. Iglesias} studied electronics engineering at the National University of Technology (UTN-FRM) in Mendoza, Argentina. In 2016 he received his Dr.-Ing degree in the field of instrumentation for solar physics from TU Braunschweig and the Max Planck Institute for Solar System Research in Germany. He is currently a math professor at UTN-FRM and a post-doctoral fellow, working on instrumentation and solar physics, at the national research council (CONICET) in Argentina.

\vspace{2ex}\noindent\textbf{Alex Feller} has studied physics at the Federal Institute of Technology (ETH) Zurich. In 2007 he received his PhD in the field of instrumentation for solar physics. He is currently working as an instrument scientist in the department Sun and Heliosphere of the Max Planck Institute for Solar System Research.

\listoffigures
\listoftables

\end{spacing}
\end{document}